\documentstyle [psfig,subfigure,amssymb]{mn}
\newif\ifAMStwofonts
\newcommand{\am}[2]{$#1'\,\hspace{-1.7mm}.\hspace{.0mm}#2$}
\newcommand{\HI}{\mbox{H\,{\sc i}}}

\newcommand{\Jykms}{\mbox{Jy~km~s$^{-1}$}}
\newcommand{\kms}{\mbox{km\,s$^{-1}$}}

\newcommand{\MHI}{\mbox{${M}_{HI}$}}
\newcommand{\MHILB}{\mbox{$M_{HI}/L_B$}}
\newcommand{\Msun}{\mbox{${M}_\odot$}}

\newcommand{\MsunLsunBfr}{\mbox{$\frac{{M}_{\odot}}{L_{\odot,B}}$}}

\newcommand{\JYKMS}{\mbox{$\frac{Jy km}{s}$}}
\newcommand{\sqarcsec}{arcsec$^{2}$}

\ifoldfss
  \ifCUPmtlplainloaded \else
    \NewTextAlphabet{textbfit} {cmbxti10} {}
    \NewTextAlphabet{textbfss} {cmssbx10} {}
    \NewMathAlphabet{mathbfit} {cmbxti10} {} % for math mode
    \NewMathAlphabet{mathbfss} {cmssbx10} {} %  "   "    "
  \fi
  \ifAMStwofonts
    \ifCUPmtlplainloaded \else
      \NewSymbolFont{upmath} {eurm10}
      \NewSymbolFont{AMSa} {msam10}
      \NewMathSymbol{\upi}     {0}{upmath}{19}
      \NewMathSymbol{\umu}     {0}{upmath}{16}
      \NewMathSymbol{\upartial}{0}{upmath}{40}
      \NewMathSymbol{\leqslant}{3}{AMSa}{36}
      \NewMathSymbol{\geqslant}{3}{AMSa}{3E}

      \let\leq=\leqslant \let\le=\leqslant
      \let\geq=\geqslant 
    \fi
  \fi
\fi % End of OFSS

\ifnfssone
  \newmathalphabet{\mathit}
  \addtoversion{normal}{\mathit}{cmr}{m}{it}
  \addtoversion{bold}{\mathit}{cmr}{bx}{it}
  \newmathalphabet{\mathbfit} % math mode version of \textbfit{..}
  \addtoversion{normal}{\mathbfit}{cmr}{bx}{it}
  \addtoversion{bold}{\mathbfit}{cmr}{bx}{it}
  \newmathalphabet{\mathbfss} % math mode version of \textbfss{..}
  \addtoversion{normal}{\mathbfss}{cmss}{bx}{n}
  \addtoversion{bold}{\mathbfss}{cmss}{bx}{n}
  \ifAMStwofonts
    \ifCUPmtlplainloaded \else
      \UseAMStwoboldmath
      \makeatletter
      \new@mathgroup\upmath@group
      \define@mathgroup\mv@normal\upmath@group{eur}{m}{n}
      \define@mathgroup\mv@bold\upmath@group{eur}{b}{n}
      \edef\UPM{\hexnumber\upmath@group}
      \new@mathgroup\amsa@group
      \define@mathgroup\mv@normal\amsa@group{msa}{m}{n}
      \define@mathgroup\mv@bold\amsa@group{msa}{m}{n}
      \edef\AMSa{\hexnumber\amsa@group}
      \makeatother
      \mathchardef\upi="0\UPM19
      \mathchardef\umu="0\UPM16
      \mathchardef\upartial="0\UPM40
      \mathchardef\leqslant="3\AMSa36
      \mathchardef\geqslant="3\AMSa3E

      \let\leq=\leqslant \let\le=\leqslant
      \let\geq=\geqslant 
    \fi
  \fi
\fi % End of NFSS release 1

\ifnfsstwo
  \DeclareMathAlphabet{\mathbfit}{OT1}{cmr}{bx}{it}
  \SetMathAlphabet\mathbfit{bold}{OT1}{cmr}{bx}{it}
  \DeclareMathAlphabet{\mathbfss}{OT1}{cmss}{bx}{n}
  \SetMathAlphabet\mathbfss{bold}{OT1}{cmss}{bx}{n}
  \ifAMStwofonts
    \ifCUPmtlplainloaded \else
      \DeclareSymbolFont{UPM}{U}{eur}{m}{n}
      \SetSymbolFont{UPM}{bold}{U}{eur}{b}{n}
      \DeclareSymbolFont{AMSa}{U}{msa}{m}{n}
      \DeclareMathSymbol{\upi}{0}{UPM}{"19}
      \DeclareMathSymbol{\umu}{0}{UPM}{"16}
      \DeclareMathSymbol{\upartial}{0}{UPM}{"40}
      \DeclareMathSymbol{\leqslant}{3}{AMSa}{"36}
      \DeclareMathSymbol{\geqslant}{3}{AMSa}{"3E}

      \let\leq=\leqslant \let\le=\leqslant
      \let\geq=\geqslant 
    \fi
  \fi
\fi % End of NFSS release 2

\ifCUPmtlplainloaded \else
  \ifAMStwofonts \else % If no AMS fonts
    \def\upi{\pi}
    \def\umu{\mu}
    \def\upartial{\partial}
  \fi
\fi

\title[The dwarf LSB galaxy population of the Virgo Cluster II.]
{The dwarf LSB galaxy population of the Virgo Cluster II. \newline
Colours and H{\Large \textbf I} line observations.}
\author[S. Sabatini et al.]
       {S. Sabatini$^1,2$, J. Davies$^1$, W. van Driel$^3$, M. Baes$^{1,4}$ \thanks{Postodoctoral
       Fellow of the Fund for Scientific Research, Flanders Belgium (FWO-Vlaanderen)}, S. Roberts$^1$,
           R. Smith$^1$, \newauthor S. Linder$^1$, K. O'Neil$^5$ \\
        $^1$ Department of Physics and Astronomy, Cardiff University, Queen's
                Building, PO Box 913, Cardiff CF24 3YB, UK \\
        $^2$ INAF-OAR, via di Frascati 33, 00040 Monteporzio Catone, Roma, Italy \\
        $^3$ Observatoire de Paris, GEPI, CNRS UMR 8111 and Universit\'e Paris 7,
              5 place Jules Janssen, F-92195 Meudon Cedex, France\\
        $^4$Sterrenkundig Observatorium, Gent, Belgium \\
    $^5$ NRAO, P.O. Box 2, Green Bank, WV 24944, U.S.A.}
\date{Draft version: 04/6/03}
\pagerange{\pageref{firstpage}--\pageref{lastpage}}
\pubyear{2002}

\begin{document}

\maketitle

\label{firstpage}

%%%%%%%%%%%%%%%%%%%%%%%%%%  ABSTRACT %%%%%%%%%%%%%%%%%%%%%%%%%%%%%%%%%%%%%%%%%%%%%%%%%%%%%%%%%

\begin{abstract}
%Dwarfs are the most common galaxies in the nearby universe, they
%may be the first galaxies formed and their environment dependent
%evolution constitutes a central problem in current astronomy.
In order to investigate the nature of dwarf Low Surface Brightness
(LSB) galaxies we have undertaken a deep B and I band CCD survey
of a 14 sq degree strip in the Virgo Cluster and applied a Fourier
convolution technique to explore its dwarf galaxy population down
to a central surface brightness of $\sim$ 26 B mag/ arcsec$^{2}$
and a total absolute B mag of $\sim$ -10. In this paper we carry
out an analysis of their morphology, (B-I) colours and atomic
hydrogen content. We compare these properties with those of dwarf
galaxies in other environments to try and assess how the cluster
environment has influenced their evolution. Field dwarfs are
generally of a more irregular morphology, are bluer and contain
relatively more gas. We assess the importance that various
physical processes have on the evolution of cluster dwarf galaxies
(ram pressure stripping, tidal interactions, supernova driven gas
loss). We suggest that enhanced star formation triggered by tidal
interactions is the major reason for the very different general
properties of cluster dwarfs: they have undergone accelerated
evolution.
\end{abstract}


%%%%%%%%%%%%% INTRODUCTION %%%%%%%%%%%%%%%%%%%%%%%%%%%%%%%%%%%%%%%%%%%%%%%%

\section{Introduction} \label{sec:intro}

%Dwarfs are the most common galaxies in the nearby universe
%(Ferguson \& Binggeli, 1994), may be the first galaxies formed
%(Blumenthal at al. 1984; White \& Frenk, 1991) and their
%environment dependent evolution constitutes an important problem
%in current astronomy.
The effect of the environment upon bright galaxies has been well
studied over the years but can still be controversial (see for
example the E-to-S0 ratio controversy: Dressler et al, 1997;
Andreon, 1998; Lubin et al, 1998). There are noticeable variations
in the morphology, colour and magnitude of cluster bright galaxies
with respect to the properties of field galaxies. A galaxy in a
cluster can be subject to many different processes that are not at
work (or less likely to occur) in the field: direct collisions,
galaxy-galaxy or galaxy-cluster tidal interactions, high/low-speed
encounters between galaxies, ram pressure stripping by the ICM,
pressure confinement and combinations of the above. The two main
effects produced by these various processes are in some ways
opposite: if, on one hand, a cluster can diminish the gas content
(and thus the star formation rate, SFR) of its galaxies by means
of various stripping mechanisms, on the other hand it can also
trigger star formation and accelerate evolution by means of tidal
interactions. Which of these two effects is prominent remains
controversial (Davies \& Phillips, 1989; Hashimoto et al, 1998;
Gnedin, 2003) and possibly depends on the exact nature of the
environment and on the galaxy type considered. Different studies
come to different conclusions: whereas some report the quenching
of star formation in clusters rather than its enhancement
(Kennicut 1983; Dressler, Thompson \& Shectman, 1985; Poggianti et
al, 1999; Dressler et al, 1999), others find the opposite result,
suggesting a similar or higher SFR in clusters than in the field
(see Bothun \& Dressler (1986) and Caldwell et al. (1993) for the
Coma Cluster). \\
Also, observations of the distant 'faint blue galaxies' have been
used to infer that star formation is enhanced (or accelerated) as
galaxies initially fall into a cluster, but at the current time
star formation is suppressed compared to the field.

These issues are further complicated when investigating the
effects of the environment upon dwarf galaxies, due to their low
magnitude and surface brightness values. If dwarfs are the first
objects formed, as predicted by Cold Dark Matter (CDM) models of
hierarchical structure formation (White \& Rees 1978; White \&
Frenk, 1991), we should expect them to be the oldest galaxies in
the universe and to be present in all environments. They should
have similar properties everywhere, unless the environmentally
dependent evolution is strong.
%The standard Cold Dark matter (CDM) picture of hierarchical
%structure formation ( . If the stellar mass of a dark halo is proportional to its
%dark matter mass, then the Luminosity Function (LF) is a direct
%measure of the dark matter Mass Function. The observed faint-end
%slope of the luminosity function when averaged over all
%environments is about -1.2 (Norberg et al. 2002); this has to be
%compared to standard CDM model predictions which typically predict
%a slope of -1.8 to -2 for the dark matter mass function
%(Kauffmann, White \& Guiderdoni 1993).
The latest observational results, however, indicate that the
faint-end-slope of the Luminosity Function (which quantifies the
number of dwarfs) has a strong environmental dependence (Trentham
\& Tully 2002; Roberts et al, 2004). If the stellar mass of a dark
halo is proportional to its dark matter mass, then the Luminosity
Function (LF) is a direct measure of the dark matter Mass Function
and model predictions require many more small dark matter halos
around individual galaxies and in galaxy clusters than have been
detected (Kauffmann, White \& Guiderdoni 1993; Moore, Lake \& Katz
1998). This failure of the standard CDM prediction is not
universal: there are galaxy clusters, like Coma, Virgo and Fornax,
where the faint-end-slope of the Luminosity Function is found to
be steep (Kambas et al. 2000; Milne \& Pritchet 2002; Sabatini et
al. 2003). Compared to these clusters, there is a real lack of low
luminosity galaxies in lower density environments, like the Local
Group and the field (Pritchet \& van der Bergh, 1999; Norberg at
al. 2002; Roberts et al, 2004). These results imply that there
cannot be a global dwarf galaxy formation suppression mechanism as is
invoked in many simulations. This normally takes the form of gas
loss through supernova driven winds (a 'feedback' mechanism).

Ignoring the possibility of many dark halos with
no baryons at all, there are two possible interpretations consistent with the observations: \\
1) CDM predictions are correct, but we need to invoke some
mechanisms that preserve primordial dwarf galaxies in some
environments, while destroying them in others; \\
2) the dwarf galaxies found in rich clusters are a different
population from the primordial one predicted by CDM - meaning that
some processes must have actually formed them solely in some
environments and/or suppressed them in others.

The most popular environment dependent mechanisms proposed in
support
of the former hypothesis are: \\
%a) Feedback by Supernovae (Dekel \& Silk, 1986) - SN driven winds
%of the first population of formed stars could expel a major part
%of gas from small CDM halos, thereby preventing further star
%formation. \\
a) Squelching (Tully at al. 2002) - a suppression of dwarf galaxy
formation in low density environments due to photoionization
occurring at a critical phase in the structure formation. Note,
however, that WMAP results now place the ionisation epoch at
z$\approx$ 20 (Spergel et al, 2003) and fluctuations of order
$10^7$ \Msun\ in CDM models are extremely rare at that epoch
(Miralda-Escude, 2003). Also, the star formation histories of
Local Group satellites show evidence of several (also recent) star
formation bursts, some continuing to the present day. For these
reasons we will not consider this process further. \\
b) Pressure confinement (Babul \& Rees, 1992) - the pressure of
the intra-cluster medium (ICM) reduces the gas loss produced by a
feedback mechanism such as supernova driven winds. \\
%c) Ram pressure stripping (Gunn \& Gott, 1972) - the pressure of
%the ICM strips the galaxy's gas away as it moves through the cluster. This process will
%be discussed in detail in Sec \ref{sec:rampressure}. \\
On the other hand, in support of the second hypothesis, mechanisms
that can form dwarf galaxies after the CDM dwarf galaxy formation epoch are: \\
c) Tidal interactions (Okazaki \& Taniguchi, 2000) - galaxy-galaxy
interactions
and mergers can result in the formation of so-called tidal dwarfs. \\
d) Harassment (Moore et al, 1999) - infalling LSB disc galaxies in
a cluster are subject to many high speed encounters that can
result in their morphological transformation into dwarf
Ellipticals (dE) galaxies.

Trying to disentangle these issues requires a better understanding
of how the mechanisms involved in galaxy evolution relate to the
environment, and thus a detailed study of the properties of the
dwarfs found in different environments: in what follows we will
try to analyse what makes a cluster different from the field or
from a loose group, and which observables (e.g., galaxy colours,
\HI\ content, velocity dispersion) can help distinguishing between
the two different interpretations stated above.

Being the nearest cluster (d $\sim$ 16 Mpc; Graham et al, 1999;
Jerjen, Binggeli \& Barazza, 2003) with several hundreds of bright
galaxies ($\sim$ 1277 sure members are listed in the Virgo Cluster
Catalogue (VCC); Binggeli, Sandage \& Tarenghi, 1984), the Virgo
cluster offers the best opportunity for the detailed study of
large numbers of bright galaxies and faint dwarfs in the cluster
environment. The most complete optical survey of the Virgo cluster
to date, the VCC, compiled using photographic plates and a visual
detection method, is complete for objects of moderate surface
brightness with $M_{B}<-14$. At present, however, there is no
published catalogue of candidate dwarf galaxy members of the
entire Virgo cluster which is complete for objects fainter than
$M_{B}=-14$ and central surface brightness values as low as 26 B
mag/\sqarcsec. These are the properties of the numerous low
luminosity galaxies dominating the numbers in the Local Group
(Mateo, 1998).

There are several studies in the literature that point to evolving
populations of dwarfs in clusters of galaxies (see Conselice et al
2001, 2003; Poggianti et al, 2001; Jerjen et al, 2004; Rakos \&
Schombert 2004), but none of them reaches the faint magnitude
values of Local Group dwarfs. In order to investigate the
possibility that previous surveys have missed a fainter population
of  Virgo Cluster dwarfs, we have applied (Sabatini et al, 2003;
paper I) an optimised Fourier convolution technique to deep CCD
images of one strip in the cluster, detecting LSB dwarfs in the
$B$ band down to surface brightness values of 26 mag/arcsec$^{2}$
and absolute magnitudes of -10 at the assumed Virgo cluster
distance. A detailed description of the selection procedure and a
determination of the Luminosity Function of the cluster dwarfs is
given in Paper I. Over a search area of $\sim$ 14 deg$^{2}$ we
have identified 231 dwarf low surface brightness candidates, 105
of them being previously uncatalogued galaxies (see VCC; Impey \&
Bothun 1988; Trentham \& Hodgkin 2002).

With all this in mind, and with the aim of improving our
understanding of the nature of very faint dwarf galaxies in the
cluster environment, we have supplemented our deep B band data
with I band data and \HI\ line follow up observations. In a
separate paper (Roberts et al, 2004) we compare our results with
those from an identical dataset (same detector, telescope,
exposure time, detection method) of both the Ursa Major Cluster
and the general field.

The paper is organized as follows: in sections \ref{sec:data} and
\ref{sec:photometry} we describe the data and the data reduction;
in section \ref{sec:DGRsec} we discuss the spatial distribution of
the dwarf galaxies and its relation with the giants; in
\ref{sec:colours} we present the B-I colours of the galaxies in
our sample and compare these results with different environments
of the Local Universe;  in section \ref{sec:hiobs} we describe the
Arecibo 21cm \HI\ line observations, data reduction and results of
the sub-sample observed; in sections \ref{sec:discussion} we
discuss the possible mechanisms occurring in the cluster
environment and their relation with our results; in section
\ref{sec:conclusions} we give our conclusions.

%%%%%%%%%%%%%%%%%%%%%%%%%% SAMPLE & B and I  PHOTOMETRY %%%%%%%%%%%%%%%%%%%%%%%%%%
\section{The Data}\label{sec:data}

The optical data that we use in this paper are part of the 2.5m
Isaac Newton Telescope Wide Field Camera (INT WFC) survey of the
Virgo Cluster (see paper I for further details). Here we present
results from the west-east B and i band strip in the cluster,
extending from the centre of the cluster (identified as M87)
eastwards for 7$^\circ$, with a total area of $\sim $14 deg$^2$.
This region of the cluster is part of the dynamical unit named
cluster A (M87), excluding members of other units like cluster B,
clouds M and W.

The data in both bands were preprocessed and fully reduced using
the Wide Field Survey pipeline: this includes de-biasing, bad
pixel replacement, non-linearity correction, flat-fielding,
defringing (for i band) and gain correction. The photometric
calibration makes use of several (5-10 per night) standard stars
and the zero-points are accurate to 1-2 per cent. For more
details, see http://www.ast.cam.ac.uk/~wfcsurv/pipeline.html. The
results throughout the paper are given in the B and I magnitudes
of the Johnson/Cousin photometric system and the conversion from
the INT colours was done using instructions given at
http://www.ast.cam.ac.uk/~wfcsurv/colours.php.

\begin{figure}
\psfig{file=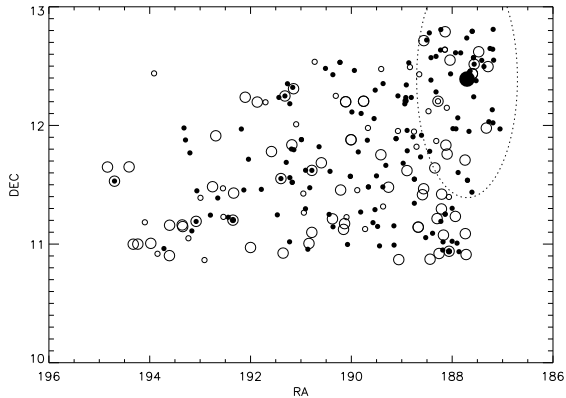,width=8cm} \caption{Detections in the
strip. Different symbols refer to different morphologies: filled
circles for dE, small open circles for dI and larger open circles
for VLSB, a circle around a dot indicates an undecided dE/dI. The
large filled dot near the top right corner is M87. We plotted a 1
degree radius circle around M87 to show the area where our \HI\
observations were too highly contaminated by this 220 Jy continuum
source. The R.A. and Dec. axes are in decimal degrees; the aspect
ratio of the field is stretched in the N-S direction - see the
plotted circle.} \label{fig:radec_detplot}
\end{figure}

In paper I we presented a convolution technique to detect and also
measure the photometric properties of faint LSB galaxies. By
applying it to the B band data, we obtained a final catalogue of
231 dwarf galaxy candidates in the strip (Fig
\ref{fig:radec_detplot}).
The use of this automated technique on homogeneous deep data
allows us to study the population of Virgo Cluster dwarf galaxies
down to the very faint limiting central surface brightness values
($\sim 26$ B mag/arcsec$^{2}$) and absolute magnitudes ($M_B \sim
-10$) typical of Local Group galaxies.

\section{Data reduction} \label{sec:photometry}

For the analysis of (B-I) colours, we need consistent measurements
of the magnitudes in both bands. Even when fully reduced, the i
band images are still affected by a fringing signature. This makes
it impossible to apply our convolution technique to measure fluxes
in this band. The data analysis for this section was therefore
performed using the aperture photometry routine from the GAIA
package on both the i and B band data for the sake of consistency.
Photometry was done using for each galaxy an aperture large enough
to include its total flux and yet avoiding contamination by nearby
objects. For each object the aperture and centre position
determined in the B band was subsequentely used in the i band as
well.

Although morphological classification is very difficult for some
of the objects, due to their very low surface brightness, we
classified the galaxies of our sample either as dE, dI or VLSB. We
describe as dE the spherical diffuse ones and the more compact
elliptical objects; dE,N denotes objects in this category that
show a nucleus. The dIs are the irregularly shaped objects, often
showing clumps. The objects that could not be assigned to either
of these category we refer to as VLSB (very low surface
brightness). The overall sample is dominated by dE types (54$\%$,
a quarter of which have a nucleus), whereas the dIs represent
 27$\%$ and the VLSBs 14$\%$.

B-I colours we measured for practically all the galaxies, with the
exception of 6: 3 have very low signal-to-noise ratios on the B
band images and are not visible at all on the i band images; for
the other 3 the i band image is corrupted. For the errors on the
B-I colours we refer to Sec \ref{sec:colours} and for more
detailed comments on single objects to Sec \ref{sec:hiresults}. \\
The total B magnitudes and central surface brightness values used
in the plots were all calculated using our convolution algorithm
(see paper I for estimated errors).

Table \ref{table:dettab} lists some of the global optical and \HI\
properties of the dwarf LSB galaxies of our sample that have been
detected in \HI\ : centre position, morphological classification,
total apparent magnitude, central surface brightness and
scale-length in B, B-I colour, rms noise of the \HI\ spectra,
integrated \HI\ line flux, $I_{HI}$, the $W_{50}$ and $W_{20}$
line widths, measured at respectively, the 50\% and 20\% level of
peak maximum, the centre velocities, $V_{HI}$, \MHI\ and the
relative \HI\ content, \MHILB (see Sec. \ref{sec:hiobs} for
details on the \HI\ data).

%Also, included in Table 3 are 15 LSB objects listed in the
%literature (Trentham \& Hodgkin (2002); VCC) as candidate Virgo
%Cluster members, that are located in the E-W strip, but were not
%picked out by our detection algorithm (for discussion on these see
%paper I). We observed these 15 sources in \HI\ (see Sec.
%\ref{sec:hiobs}), as most were classified as dI and therefore
%potentially \HI\ gas-rich.

\section{The Dwarf-to-Giant Ratio}\label{sec:DGRsec}

In paper I we have shown the radial profile of the galaxy number
density of our sample. This can be compared with that of the
bright galaxies. We have defined a Dwarf-to-Giant Ratio (DGR) as
the ratio of dwarf galaxies (defined as those with $-14 \leq M_{B}
\leq -10$) to giant galaxies (with $M_{B} \leq -19$ \footnote{This
definition for the giants is assumed in order to easily compare
our numbers with the ones in the Local Group, as this is the
magnitude limit in order to include its three giants M31, M33 and
the Milky Way.}). This is a simpler and more versatile measure to
quantify the Luminosity Distribution than the LF. We have shown
that the DGR remains rather flat with distance from M87 with a
median value of $\sim 20$. This number has to be compared with a
value of about 6, for the field population (Roberts et al, 2004).

It is difficult to compare our results with other studies, as
different papers use different definitions for the DGR. Ferguson
\& Sandage (1991) studied 7 groups and clusters and defined an
elliptical giants-to-dwarf-ratio EDGR. Our results are consistent
with the strong correlation found by these authors of the EDGR
with the cluster richness: they found as well that the EDGR in the
Virgo Cluster appears to be independent of distance from the
cluster centre. Secker \& Harris (1996), however, find an EDGR in
the Coma Cluster identical to the one in the less rich Virgo
Cluster and thus in contrast with the EDGR-richness correlation.
They point out that the presence of substructure could be an
important factor and that the Coma Cluster result would be
consistent with the cluster being built up from the mergers of
less rich clusters. Also, more recently, Driver et al (1998)
studied 7 rich clusters at redshift $\sim 0.15$ and found an
anti-correlation between richness (or density) of the cluster and
DRG; their definition of dwarfs, however, comprise all galaxies
fainter than -19.5 and it is therefore difficult to compare their
results with ours.

Our results have also to be compared to the values in the Local
Group: in a hierarchical structure formation scenario, clusters
like Virgo, should be formed by accretion of smaller groups. It is
thus possible that some or all the dEs in Virgo were once part of
structures like the Local Group that fell into Virgo with their
spirals and satellites. %In order to investigate this aspect, it is
%important to measure the number and spatial distribution of dwarfs
%in groups and in the field and compare it with clusters like
%Virgo.
The {\it expected} DGR from hierarchical structure formations,
should be the same for all these environments, or a smaller one in
the cluster if the disruptive processes are the dominant ones for
dwarf galaxies, as often assumed in $\Lambda$CDM models of galaxy
formation. Contrary to these predictions, the {\it observed}
increase in the DGR that we find  in clusters would instead imply
that some galaxies are actually 'formed' in the assembling of the
cluster. \\ %A complete discussion on the possible processes at
%work in clusers and their influence on the dwarf population is
%given in section \ref{sec:discussion}. \\
\begin{figure}
\psfig{file=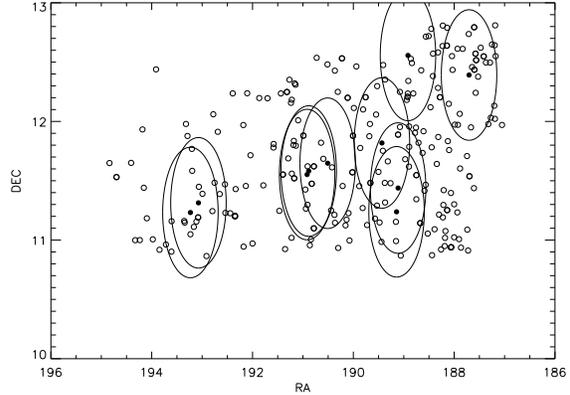,width=8cm}
\caption{\footnotesize{Spatial distribution of the dwarf galaxies
of our sample (open circles) and of the giant galaxies of the
Virgo Cluster (dots), defined as the ones with $M_{B} \le -19$.
Large circles are the tidal radii of the giant galaxy that is at
their centre.}}\label{fig:DGR_cerchi}
\end{figure}
Dwarfs in the Local Group are mainly associated with bright
galaxies: 75\% of them cluster in 3 subgroups respectively around
M31, the MW and NGC3109 and the other 25\% are part of the Local
Group Cloud, populated only by dwarf irregular systems (Mateo
1998). It is interesting to compare these properties with the
spatial distribution that we find in the Virgo Cluster. Fig.
\ref{fig:DGR_cerchi} shows the spatial distribution of our sample
dwarfs and of the Virgo Cluster giants (M$_B \leq -19$) from the
VCC. We have also plotted the tidal radius around each giant
galaxy in the strip (larger circles). This is the truncation
radius of the dark matter halos in the cluster and it can
therefore be considered as the one within which sub-halos remain
bound to the galaxy potential well. Describing the cluster and the
galaxy halos with a singular isothermal distribution \footnote{
Whether the Virgo Cluster is relaxed is still a matter of debate
(see Binggeli et al 1987; Binggeli et al 1993; Schindler et al
1999); nevertheless
 substructures in the cluster will amplify the truncation
effect (Gnedin 2003) and would therefore reduce the radii of the
circles in Fig \ref{fig:DGR_cerchi}.}, the tidal radius $R_{t}$
can be approximated as:

\begin{equation}\label{eq:truncradbright}
R_{t}\approx \frac{\sigma_{g}}{\sigma_{cl}}R_{p}
\end{equation}

where $R_{p}$ is the distance of the closest approach to the
centre of the cluster, and $\sigma_{g}$ and $\sigma_{cl}$ are
respectively the velocity dispersions of the galaxy and of the
cluster (Merritt 1984; Gnedin, 2003). For many galaxies the
distance of closest approach to the cluster centre is of order the
cluster core radius (Gnedin 2003) and for the Virgo Cluster we can
therefore approximate it with 0.5 Mpc (Binggeli, Tammann \&
Sandage, 1987). Assuming the velocity dispersion of the galaxies
in the cluster to be $\sim$700 km/s and the rotational velocity of
a giant $\sim 200$ km/s, the typical tidal radius for a giant
galaxy in the Virgo cluster is $R_{t}\sim 150$ Kpc.

Although the spatial distribution in Fig \ref{fig:DGR_cerchi} is
just a projection and it is difficult to know what the real
distribution is, it is interesting that a significant part of the
dwarf galaxy population does not seem to be associated with the
giants. A similar result was found by Ferguson (1992): in his
analysis of bound companions in the Virgo Cluster, he suggested
the existence of a free-floating cluster member population made of
stripped companions. In agreement with this view, a more recent
paper by Conselice et al (2001) showed that there is little
evidence for a dynamically cool dE component, as might be expected
if a significant fraction of dE were bound to individual cluster
galaxies (see also Binggeli, Popescu \& Tammann, 1993; Binggeli,
Tammann \& Sandage, 1987). Our results indicate as well that there
appears to be a cluster dwarf population: $\sim 40\%$ of the
galaxies in our sample are apparently not bound to the giants and
this is a larger number than that in the Local Group. Again this
number cannot be accounted for if the Virgo Cluster was simply
made up of groups like the Local Group. Also, after subtracting
the average number density of this 'un-bound' dwarf population
from the average number of dwarf galaxies within a projected tidal
radius from a giant, the number of dwarfs possibly associated with
each giant is $\sim 13 \pm 3$. This has to be compared with a
value of 3 for the Milky Way (this would be obtained if the Milky
Way was at the Virgo Cluster distance and for dwarf galaxies that
match our selection criteria) and 4 for the Local Group,
suggesting that infalling galaxy groups cannot supply sufficient
dwarfs in cluster (see also Conselice et al, 2001; Tully et al,
2002).

%%%%%%%%%%%%%%%%%%%%%%%%%%%%%%%%%%%%%%%%%%%%%%%%%%%%%%%%%%%%%%%%%%%%%%%%%%%%%
\section{B-I colours} \label{sec:colours}

\begin{figure}
\psfig{file=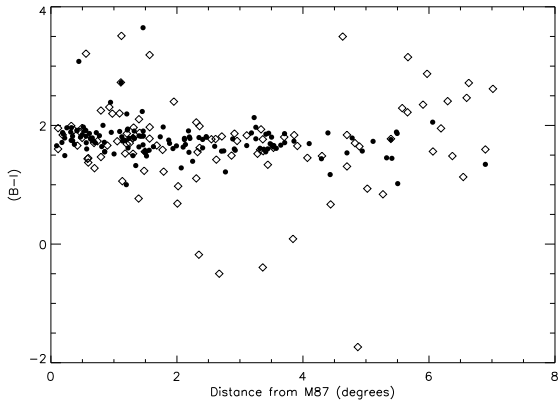,width=8cm} \caption{\footnotesize{B-I
colours plotted as function of distance from M87, assumed as the
centre of the cluster. Different symbols refer to different
morphological types: filled circles for dE and diamonds for dI and
VLSB. The sample does not seem to have any dependence upon
distance from the cluster centre. Average error on the (B-I)
colour value is 0.25. Points which deviate more from the mean
value have larger errors, as shown in Fig \ref{fig:col_mag}.}}
\label{fig:radial_bmi}
\end{figure}
\begin{figure}
\centerline{\psfig{file=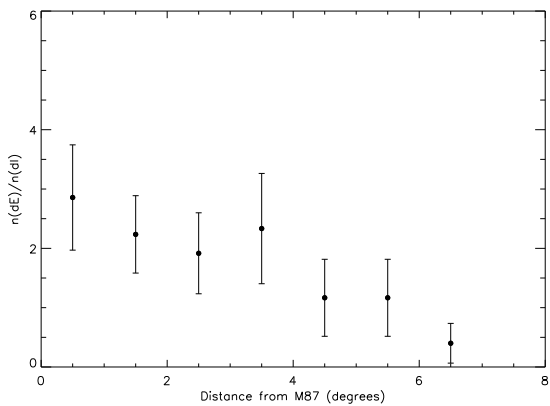,width=8cm}}
\caption{\footnotesize{dE-to-dI ratio as a function of
cluster-centric distance.}} \label{fig:ratiodEdI}
\end{figure}

Being interested in the nature of the dwarf galaxies in our sample
and in possible environmental effects on the evolution of galaxies
in the cluster, we firstly investigated the possibility of a
relationship between colours and cluster-centric distance. In Fig
\ref{fig:radial_bmi} we plot the B-I integrated colours of our
galaxies as a function of their position in the cluster
(identified as the distance from M87); different symbols refer to
different morphologies: filled dots for dEs and open diamonds for
dIs and VLSBs. The range of colours is quite wide, with an average
value of about 1.7. Even if this figure shows that the colours do
not appear to have any dependence upon distance from the centre of
the cluster, the numerical ratio of dEs to dIs (Fig
\ref{fig:ratiodEdI}) shows a morphology-density relation. This, we
believe, could be an important clue to the nature of the cluster
dwarf galaxy population: the well ordered spherically symmetric
galaxies (dE) are preferentially found towards the cluster centre,
the more irregular galaxies (dI) towards the cluster edge. This is
directly analogous to the bright galaxy morphology-density
relation (Dressler, 1984). In the same way as it has been proposed
that spiral galaxies are being transformed into S0 galaxies in the
cluster environment (Kodama \& Smail 2001 and references therein)
this could indicate that infalling dI galaxies are transformed
into dEs (Davies \& Phillipps 1989; Conselice et al 2003).

\begin{figure}
\centerline{\psfig{file=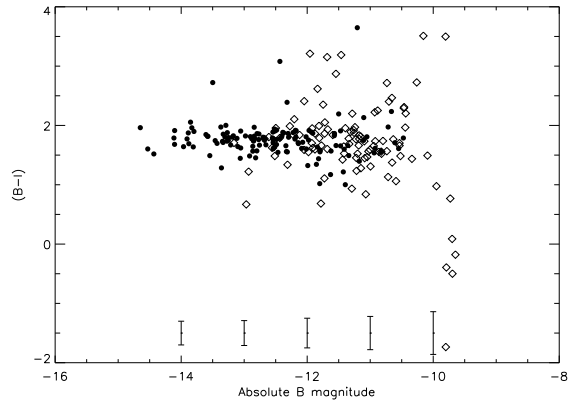,width=8cm}}
\caption{\footnotesize{B-I colours plotted as function of total B
absolute magnitude.}} \label{fig:col_mag}
\end{figure}

\begin{figure}
\centerline{\psfig{file=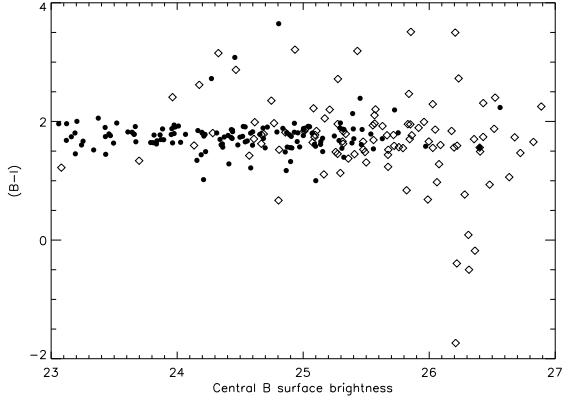,width=8cm}}
\caption{\footnotesize{B-I colours plotted as function of central
B surface brightness.The colours appear to be spread out along the
range of magnitudes considered without any evident correlation,
although it has to be noticed that the extremely blue colours are
found for $\mu_{0}< 1\sigma_{sky}$ ($ \sim $ 26 B mag/\sqarcsec).
Average error on the (B-I) colour value is 0.25. Points with large
deviation from the mean value have larger errors, as shown in Fig
\ref{fig:col_mag}.}} \label{fig:col_sb}
\end{figure}

\begin{figure}
\centerline{\psfig{file=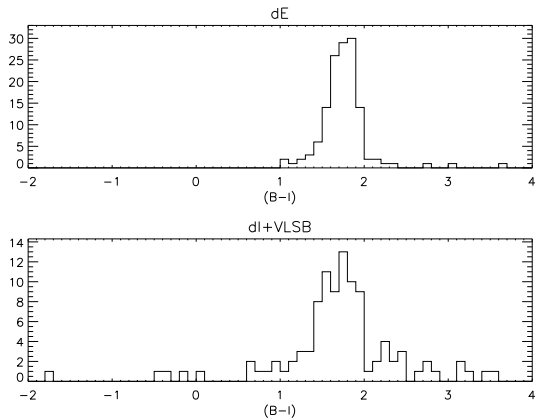,width=8cm}}
\caption{\footnotesize{(B-I) colour distribution for dE and dI
respectively. Median and standard deviation values for the 2
distributions are respectively: (B-I)$_{dE}$=1.76,
$\sigma_{dE}$=0.30 ; (B-I)$_{dI}$=1.70, $\sigma_{dI}=0.70$}.}
\label{fig:morph_histo}
\end{figure}

Figures \ref{fig:col_mag} and \ref{fig:col_sb} show the
distribution of colours as a function of total absolute magnitude
and central surface brightness, respectively. Again, the
distributions do not show any correlation. Fig \ref{fig:col_mag}
clearly shows that the scatter in colours has a dependence on
total magnitude. In this figure we also give an estimate of the
average error at each magnitude. The errors on the colours were
calculated considering the following two independent
contributions: 1) a systematic error due to the aperture
photometry procedure; 2) the error in the calculation of the sky
subtraction, that depends on the area of the object. Although the
errors on the colour increase at the fainter magnitudes, the
scatter is still much larger than the calculated errors. Fig
\ref{fig:morph_histo} shows that dE galaxies also have a narrower
range of colours across the whole magnitude range
((B-I)$_{median}$=1.76 ; $\sigma$=0.30), while dI galaxies have a
much broader range ((B-I)$_{median}$=1.70 ; $\sigma$=0.70)
possibly indicating their different evolutionary states i.e.
undergoing, fading from or between a burst of star formation. For
comparison, Karick, Drinkwater \& Gregg (2003) also find a wide
range of colours for their Fornax cluster dwarf galaxies, larger
than that of brighter galaxies.


\subsection{Comparison with synthetic colours}

\begin{figure}
\centerline{\psfig{file=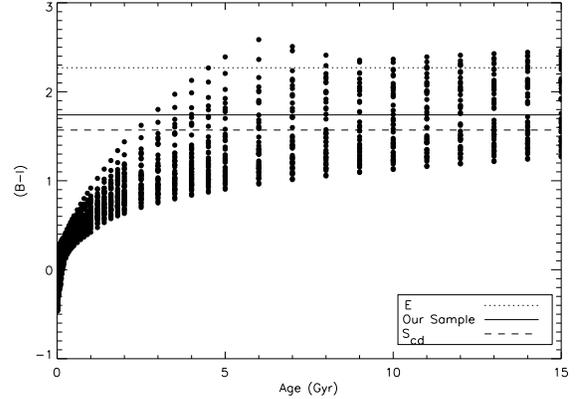,width=8cm}}
%\begin{figure}
%\psfig{file=bmi_colour_pegase.ps, width=2cm}
\caption{\footnotesize{Synthetic (B-I) colours as function of time
produced using the PEGASE evolutionary synthesis code. Overlayed
in the plot, as explained in the legend, are the median (B-I)
colour of our distribution (filled line) and for comparison a
typical value for an Elliptical Galaxy (dotted line) and for a
late-type Spiral (dashed line). These latter values are taken from
Fukugita et al, 1995}} \label{fig:pegase_bmi}
\end{figure}

For comparison with our colour distribution, in Fig.
\ref{fig:pegase_bmi} we show the (B-I) colour predictions obtained
using the PEGASE evolutionary synthesis code (Fioc \&
Rocca-Volmerange 1997). The figure shows (B-I) as a function of
time and different points
for each age refer to different initial metallicities and star
formation rates chosen in the simulation. Aware that one colour is
not enough for disentangling the effects of metallicity and age
(i.e. the age-metallicity degeneracy; Worthey 1994 and references
therein), we did not explore the synthesis code in all its
potentialities and we run it instead, using the default choices
for the many allowed parameters: a Salpeter IMF (with lower mass
0.1 \Msun\ and upper mass 120 \Msun), a range of metallicities
(0.1, 0.05, 0.02, 0.008, 0.004, 0.0004, 0.0001 Z$\odot$) and
different exponential star formation rates with decay times
1,2,4,8,16,32 Gyr.

The main intention with this comparison is to show that the
distribution of galaxy colours of our sample lies well in the
average range of the expected values from synthetic models (Fig 8,
filled line) . Also, some of our extremely blue colours are
compatible with very young ages, but one colour is not enough to
claim any definitive conclusion and, as we have shown, errors on
these extremely blue colours are quite large. As a comparison,
average values for an Elliptical galaxy and a late-type Spiral
(Fukugita et al, 1995) have been overlayed in the plot of Fig 8.
It is interesting that the average colour of the dwarfs in our
sample is closer to that of a Spiral and bluer than that of an
Elliptical, suggesting that these dwarfs have possibly consumed
their gas until recent epochs. However, Fig \ref{fig:pegase_bmi}
clearly shows that it is not possible to obtain conclusive results
from this analysis with one colour only; combined observations in
a third band (e.g. the K band) would help disentangling the
age-metallicity degeneracy. We will be doing this as part of the
near infrared survey UKIDSS - (see, www.ukidss.org). In the
following sections, however, as a working hypothesis, we will
assume that the galaxies in our sample have similar metallicity so
that the relative B-I colour differences are indicative of
differences in age.

\subsection{Comparison with different environments of the Local
Universe}\label{sec:diffenv}

\begin{figure}
\centerline{\psfig{file=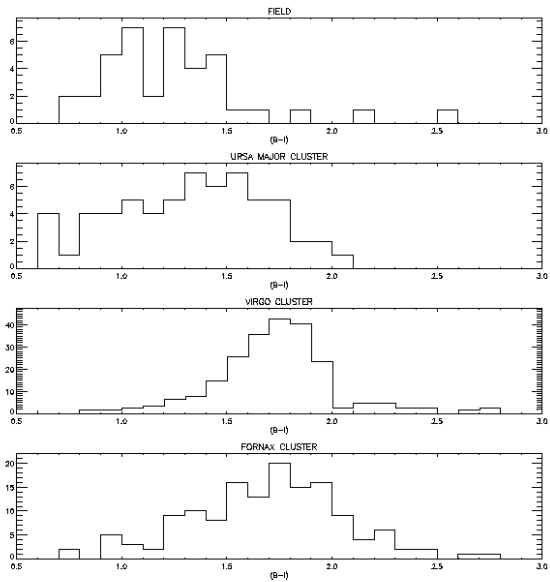,width=8cm}}
\caption{\footnotesize{B-I colour distribution of dwarf galaxies
in different environments, from top to bottom: the field, the Ursa
Major, the Virgo and the Fornax Clusters.}}
\label{fig:histo_colours}
\end{figure}

Although it is in principle difficult to combine results from
samples selected in different ways, in Fig.
\ref{fig:histo_colours} we plot the distribution of the (B-I)
colours of galaxies from our sample and for comparison the same
distributions for dwarf galaxies in different environments of the
nearby universe: the field (Makarova, 1999), the Ursa Major
Cluster (Trentham, Tully \& Verheijen 2001), the Virgo Cluster
(from our catalogue) and the Fornax Cluster (Karick, Drinkwater \&
Gregg 2003). Table \ref{table:colours} gives the median and standard
deviation for these distributions, along with some relevant
properties of the environments.

\begin{table*} %[h!]
\begin{center}
%\begin{minipage}{60mm}
%\centerline\textbf{ Slope of Virgo LF = -1.0} \centering
%{\tiny
\begin{tabular}{|c|c|c|c|c|}
\hline
Environment & Mass & Crossing time & (B-I)$_{\tiny median}$ & STDEV \\
            &(10$^{14}$ \Msun)&  (H$_{0}^{-1}$) &     & \\
\hline
nearby (field) &---&---& 1.22 & 0.36 \\
Ursa Major     &0.5&0.5& 1.38 & 0.42 \\
Virgo          &8.9&0.1& 1.74 & 0.54 \\
Fornax         &0.7&0.1& 1.83 & 0.24 \\
\hline
\end{tabular}
\caption{\footnotesize{Median and standard deviation for dwarf
galaxies B-I colours distribution in different environments for
comparison with our results in the Virgo Cluster. We also list
some of the possibly relevant properties of the environments. }}
\label{table:colours}
%}
%\end{minipage}
\end{center}
\end{table*}

Table \ref{table:colours} and figure \ref{fig:histo_colours}
suggest that the average colour distribution of dwarf galaxies
becomes progressively redder, when proceeding from the field to
denser, more elliptical galaxy-dominated environments. Gallagher
\& Hunter (1986) found the same trend comparing dI of the Virgo
Cluster to a field sample. Although the nature of the dwarfs
considered in the samples of Fig \ref{fig:histo_colours} might be
different (ours being very faint), this result seems to indicate a
strong environmental effect on the stellar population of these
galaxies: the cluster environment progressively affects the
evolution of dwarf and giant galaxies in the same way. Whether
this is due to a progressively enhanced SFR in denser environments
or to stronger gas stripping that stopped SF earlier, leaving the
cluster with an old galaxy population, needs further investigation
on the exact nature of the stellar population of these dwarfs
galaxies.
%Moreover the
%influence of the environment on dwarf galaxies is complex: on the
%one hand we can expect that their shallow potential wells makes
%them much more fragile to external perturbations than their giant
%counterparts, while on the other hand their smaller sizes make tidal
%encounters less likely to happen.
A more detailed discussion concerning these points is deferred to
Sec \ref{sec:discussion}.

\section{\HI\ observations} \label{sec:hiobs}
The \HI\ data are 21cm \HI\ line pointed observations with the
305m Arecibo telescope of a subsample of galaxies from our
catalogue. Many \HI\ surveys have been made of, or in, the Virgo
cluster area, also aimed at detecting dwarf galaxies (Conselice et
al., 2003; Hoffman et al., 1985, 1989). A blind \HI\ line survey
of the entire cluster, providing a comprehensive census of
gas-rich objects, has recently been carried out by the HIPASS
team, but the data is, as yet, unavailable.
%However, with an rms noise level of about 12 mJy this
%survey does not have the required sensitivity to detect really
%faint dwarf galaxies in the cluster: the 3$\sigma$ \HI\ mass
%detection limit for a typical 75 \kms\ wide flat-topped dwarf
%profile is about $2 \times 10^{8}$ \Msun, corresponding to a
%gas-rich ($(M_{HI}/L_{B})_{\odot} \sim $1) dwarf of absolute
%magnitude $M_B$$\sim$--16 at d=17 Mpc (----> change to 16
%Mpc!!!!).
Davies et al (2004) have carried out a deep blind Jodrell Bank
survey of a 4$\times$8 deg region in the Virgo Cluster. The
results from this survey indicate a relative lack of low mass \HI\
objects compared, to the field: galaxies in the cluster
environment are depleted in gas compared to field galaxies -
either through the gas loss mechanisms or through accelerated star
formation. This survey has also, surprisingly, shown the presence
of 2 objects without optical counter parts (Davies et al, 2004).
Although a further investigation is required in order to confirm
these objects, this result shows how \HI\ observations can still
provide new and interesting results in the Virgo Cluster.

The 3$\sigma$ \HI\ mass detection limit for a typical 75 \kms\
wide flat-topped dwarf profile for the Davies et al blind survey
is about $10^{8}$ \Msun, corresponding to a gas-rich
($(M_{HI}/L_{B})_{\odot} \sim $1) dwarf of absolute magnitude
$M_B$$\sim$-14.5 at d=16 Mpc. This is the typical value of the
brightest dwarfs in our sample. With its extraordinary sensitivity
the Arecibo telescope offers the opportunity to go deeper on
specific targets and easily reach typical \HI\ mass detection
limits at the Virgo Cluster distance of an order of magnitude
smaller (see Sec \ref{sec:hinondet}).

The general gas deficiency of Virgo Cluster members is a well
established phenomenon (Solanes et al, 2002 and references
therein; Davies et al, 2004). Deep \HI\ pointed observations of 59
Virgo cluster dEs from the VCC  by Conselice et al. (2003) yielded
only 7 clear detections (just 2 of which are new) for objects
having a mean blue absolute magnitude $M_B$ of -16 (range: -14.2
to -17.0) and with an average \MHI=2.5$\pm$3.6 10$^8$ \Msun,
$(M_{HI}/L_{B})_{\odot}=0.57 \pm $0.52 and FWHM line width
$W_{50}$=125$\pm$93 \kms. Even if the expected detection rate was
low, we carried out \HI\ follow ups of a sub-sample of our
galaxies, because dwarf galaxies in clusters do include
star-forming, and generally \HI-rich, dwarf irregulars (e.g.,
Gallagher \& Hunter 1984) as well as quiescent, gas-poor dwarf
elliptical/spheroidals (e.g., Ferguson \& Binggeli 1994; Gallagher
\& Wyse 1994). \HI\ observations are required for a clearer
understanding of their nature, in order to confirm their cluster
membership, place lower limits on the \HI\ mass of the
non-detected objects, obtain \MHILB\ ratios or upper limits and
look for environmental effects on their gas content.

\subsection{Sample description}

From our catalogue of 231 candidate LSB dwarf members of the Virgo
Cluster, we observed 100 objects in \HI\ according to the
following priorities during the observing run: we firstly selected
objects that were not part of the VCC catalogue, that didn't lie
within 1 degree from the strong continuum source M87 (see Sec.
\ref{sec:hidatared}) and we gave priority to dI or VLSB
morphological types. The galaxies observed at Arecibo are listed
in Table \ref{table:dettab} if detected in \HI\ . The galaxies are
given a name from our catalogue and a name from the literature, if
previously catalogued; the cross-correlation with other catalogues
was made within a 15$''$ radius search area around our optical
centre positions.

In addition, included in our observations are 15 LSB objects
listed in the literature (Trentham \& Hodgkin (2002); VCC) as
candidate Virgo Cluster members, that are located in the E-W
strip, but were not picked out by our detection algorithm (for
discussion on this point see paper I). We observed these 15
sources in \HI\, as most were classified as dI and therefore are
potentially \HI\ gas-rich.

%We also observed 2 of the 3 dI VCC galaxies that are in the strip
%but not in our catalogue (VCC1558, VCC1905); we did not observe
%the third, VCC 1413 (= 169), for which Hoffman (1989) reported an
%\HI\ non-detection with an rms noise of 1.0 mJy, as it is too
%close to M87 to get a proper spectrum (see Sec. 3).

%We excluded the following 3 objects from \HI\ observations at Arecibo, as they
%already had published \HI\ detections (see section 5.1 for further details):
%VCC 1889, VCC 2006 and VCC 2062.

%%%%%%%%%%%%%%%%%%%%%%% HI DATA REDUCTION %%%%%%%%%%%%%%%%%%%%%%%%%%%%%%%%%%%%%%%%%%%

\subsection{Data reduction} \label{sec:hidatared}

Data were taken with the L-Band Narrow receiver using nine-level
sampling with two of the 2048 lag subcorrelators set to each
polarization channel. All observations were taken using the
position-switching technique, with the blank sky (or OFF)
observation taken for the same length of time, and over the same
portion of the Arecibo dish as used for the on-source (ON)
observation. Each 3min+3min ON+OFF pair was followed by a 10s
ON+OFF observation of a well calibrated noise diode. The overlaps
between both subcorrelators with the same polarization allowed a
wide velocity search while ensuring an adequately coverage in
velocity. The velocity search range was -1000 to 11,000 \kms, as
the Virgo Cluster extends from 500 to 2500 \kms (Binggeli, Popescu
\& Tammann 1993).
%in practice, however,
%this range was restricted for some observations (see Table 2), due to the fact that
%2 out of the 4 correlator boards were occasionally not functioning properly.
The instrument's HPBW at 21 cm is \am{3}{6}$\times$\am{3}{5} and
the pointing accuracy is about 15$''$.
\begin{table*}
{\tiny
\begin{tabular}{l|l|c|c|l|r|r|r|r|r|r|r|r|r|r|r|r}
\multicolumn{17}{l}{{\small \textbf{\centerline{\HI\ detections from our survey}}}}\\
\hline  Obj. & Ident & R.A. & Dec. & T &  $m_{B_{T}}$ &
$\mu_{0,B}$ & $\alpha_B$ & B-I & rms & $I_{HI}$
  & $W_{50}$ & $W_{20}$ & $V_{HI}$ & d & $M_{HI}$ & \MHILB\ \\
 &  & \multicolumn{2}{c}{J2000.0} & & mag &  & ($''$) & & {\tiny (mJy)} & {\tiny (\JYKMS)} & {\tiny (km/s)} & {\tiny (km/s)}
  & {\tiny (km/s)} & {\tiny (Mpc)} & {\tiny (10$^8$ \Msun)} &  {\tiny (\MsunLsunBfr)} \\
\hline
% n1     n2       ra      dec    t     mag     csb  alfa  B-I   rms     Ihi    W50   W20    vel   dist   Mhi   M/Lb
144  &       & 123841 & 115843 &  dI  & 19.8 & 26.6 & 9 &   -  & 0.8 &  0.47 &  77 & 125 & 1036 & 16   & 0.3 & 6.2  \\
158  & TH152  & 125107 & 120339 & dI & 17.1 & 23.3 & 7 & 1.01 & 0.9 &  1.71 &  63 &  83 & 1788 & 16   & 1.0 & 1.7  \\
249  & U8061 & 125644 & 115557 &  dI  & 17.0 & 23.7 & 9 & 1.26 & 1.5 &  2.12 &  69 &  89 &  563 & 16   & 1.3 & 1.9 \\
4    & TH225  & 124112 & 105601 &    & 17.7 & 23.6 & 6 & 1.17 & 0.9 &  1.13 & 175 & 125 & 6468 & 85.5 &19.8 & 2.0  \\
27   &       & 124934 & 121411 &    & 18.2 & 23.3 & 4 & 1.04 & 1.0 &  1.03 & 197 & 297 & 7146 & 94.7 &22.0 & 3.1  \\
%121* &       & 124338 & 113720 &    & 19.9 & 25.4 & 5 & 1.93 & 0.9 &  0.59 & 160 & 172 & 1424 & 16   & --- &  --- \\
\hline
\end{tabular}
} \caption{Objects detected in \HI\ at Arecibo. [1] Object's
number from our catalogue, [2] identification from other
catalogues (T: Trentham \& Hodgkin 2002; U: UGC), [3] right
ascension (J2000.0), [4] declination (J2000.0), [5] morphological
type, [6] total apparent blue magnitude, [7] central blue surface
brightness, [8] disc scale-length, [9] B-I , [10] rms noise level
of \HI\ spectrum, [11] integrated \HI\ line flux, [12] \HI\
profile width at 50\% of peak maximum, [13] same at 20\%, [14]
\HI\ line centre velocity, [15] distance, [16] \HI\ mass and [17]
\HI\ mass to blue luminosity ratio.}\label{table:dettab}
\end{table*}

A filter was used in order to cut off all emissions at frequencies
below 1371 MHz, thus eliminating radio frequency interference
(RFI) from radars on Puerto Rico. The strongest RFI signal noted
was centred on 1381 MHz (or 8300 \kms), which however did not
occur frequently.

Using standard IDL data reduction software available at Arecibo,
corrections were applied for the variations in the gain and system
temperature with zenith angle and azimuth, a baseline of order one
to three was fitted to the data, excluding those velocity ranges
with \HI\ line emission or radio frequency interference (RFI), the
velocities were corrected to the heliocentric system (using the
optical convention) and the polarisations were averaged.  All data
were boxcar smoothed to a velocity resolution of 15 \kms\ for
further analysis. For all smoothed spectra the rms noise level was
determined and for the detected lines the central velocity, line
widths at, respectively, the 50\% and 20\% level of peak maximum,
and the integrated flux were determined.

The extremely strong continuum emission from M87 aversely affected
the rms noise level and the quality of the baselines in a sizeable
area ($\sim 1^{\circ}$) surrounding it. M87 is a 220 Jy source at
1415 Mhz, with a core and two lobes extending some 15$'$ from the
core. Attempts were made to correct the bandpass by applying a
double-switching technique in which an ON+OFF spectrum of a target
galaxy is normalised with an ON+OFF spectrum of the continuum
source influencing the data, i.e. M87, taken just after. Although
this technique has been applied successfully at Arecibo to \HI\
data of galaxies with continuum sources (e.g., Ghosh \& Salter,
2002), the situation near M87 is fundamentally different as it is
detected through the far side-lobes of the telescope, rather than
through the main beam, and it is strong enough to cause non-linear
saturation effects in the receiver system. Since we could not
obtain proper quality data within about a degree from M87, we
stopped observing objects in that area (see Fig
\ref{fig:radec_detplot}).

%\begin{figure*}
%\subfigure{\psfig{file=vc144_avg_plot.ps,width=6cm}}%,angle=-90}}
%\subfigure{\psfig{file=vc158_plot.ps,width=6cm}}%,angle=-90}}
%\subfigure{\psfig{file=vc249_avg_plot.ps,width=6cm,angle=-90}}
%\subfigure{\psfig{file=vc4_plot.ps,width=6cm,angle=-90}}
%\subfigure{\psfig{file=vc27_plot.ps,width=6cm,angle=-90}}
%\subfigure{\psfig{file=t215_plot.ps,width=6cm,angle=-90}}
%\caption
%{HI spectra for the 6 non-confused spectra detections (excluding 121)}
% \end{figure*}

\begin{figure*}
\subfigure{\psfig{file=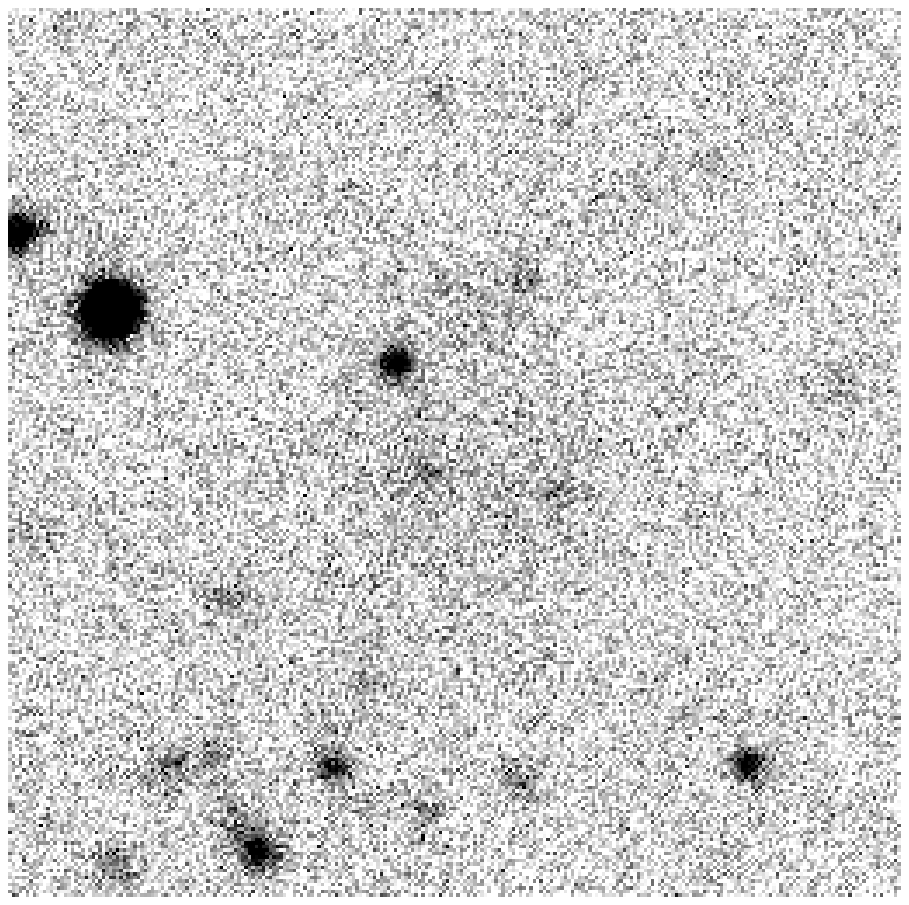,width=3.5cm}} \hspace{1cm}
\subfigure{\psfig{file=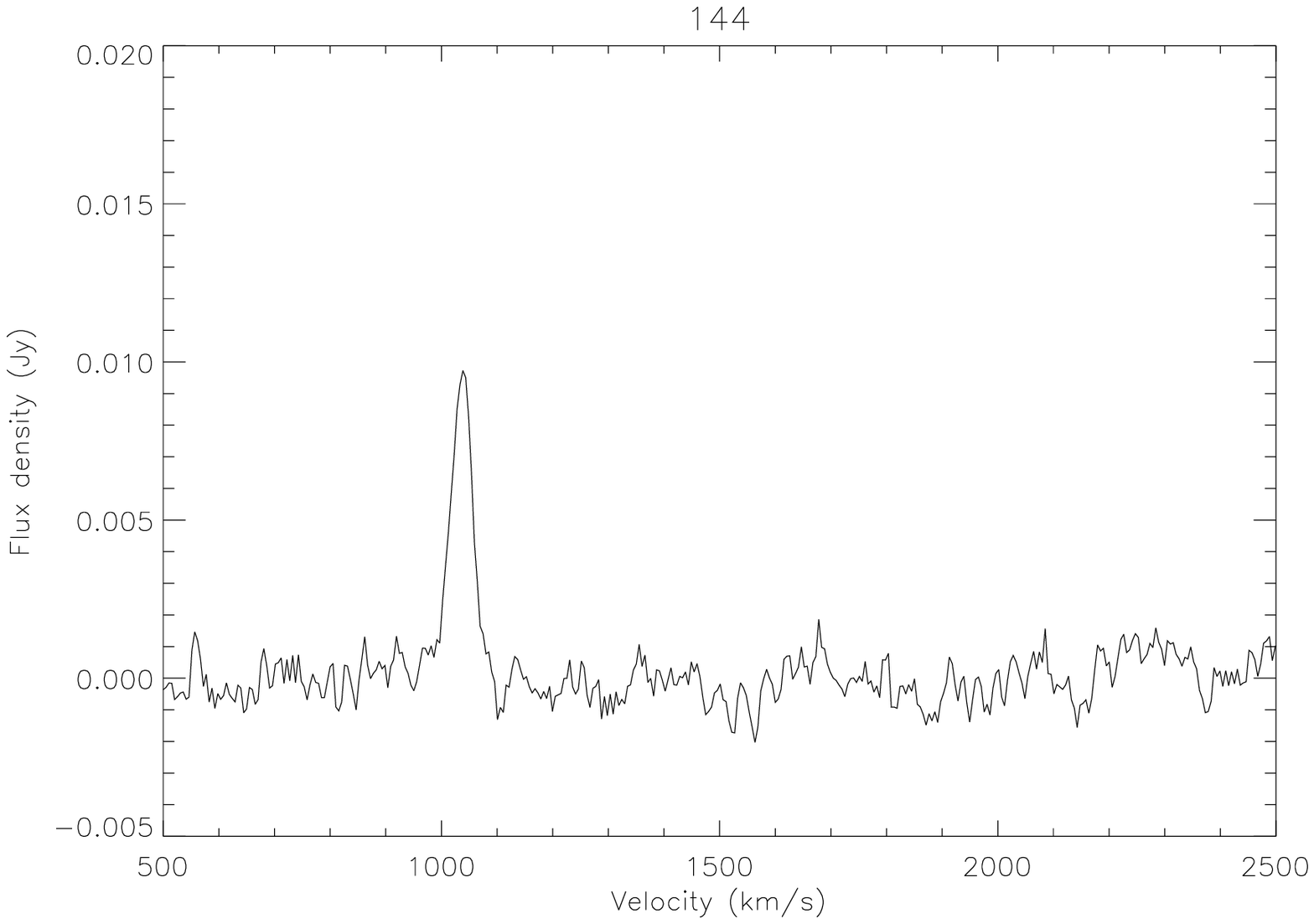,height=3.5cm}}\\
\subfigure{\psfig{file=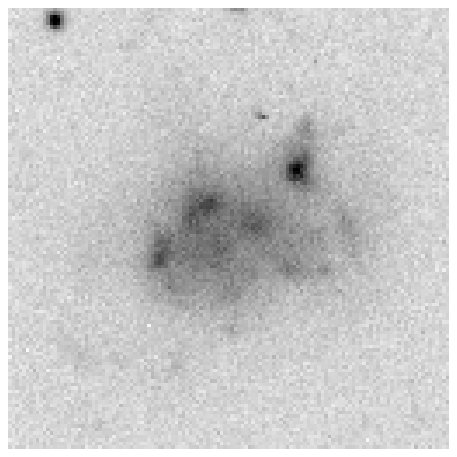,width=3.5cm}} \hspace{1cm}
\subfigure{\psfig{file=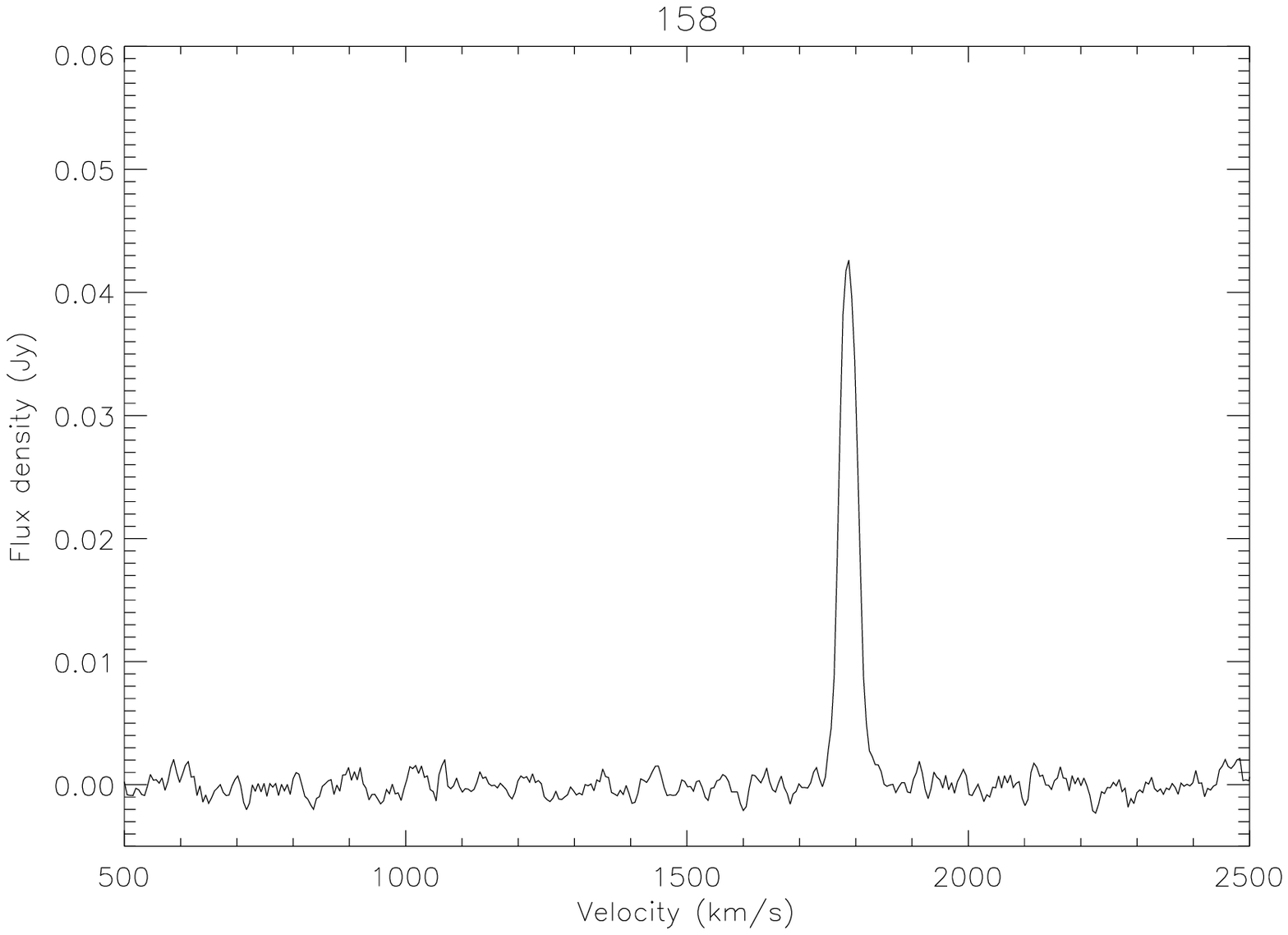,height=3.5cm}}\\
\subfigure{\psfig{file=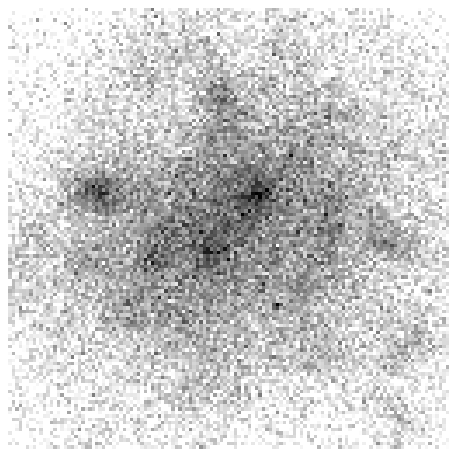,height=3.5cm}} \hspace{1cm}
\subfigure{\psfig{file=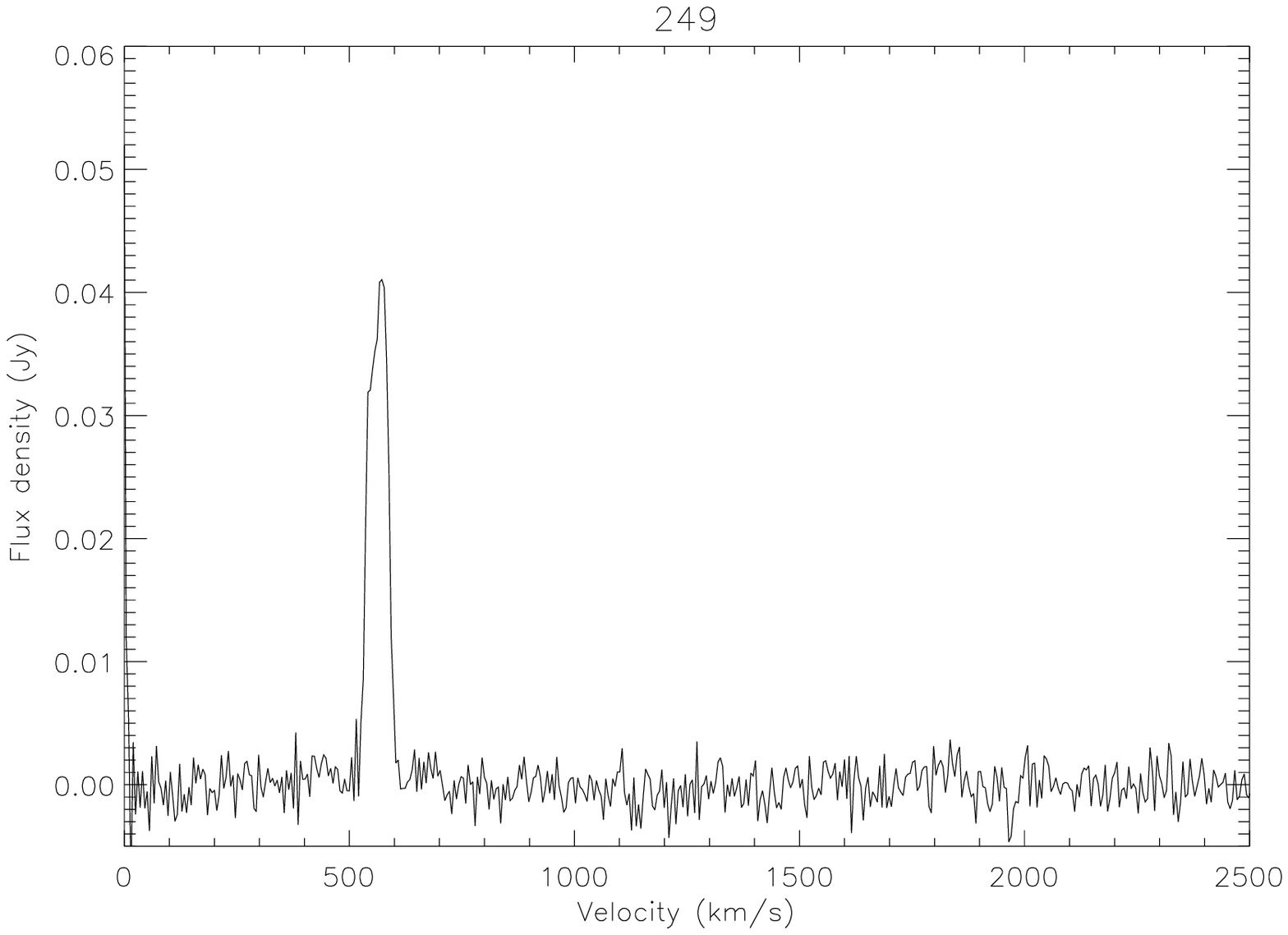,height=3.5cm}}\\
\subfigure{\psfig{file=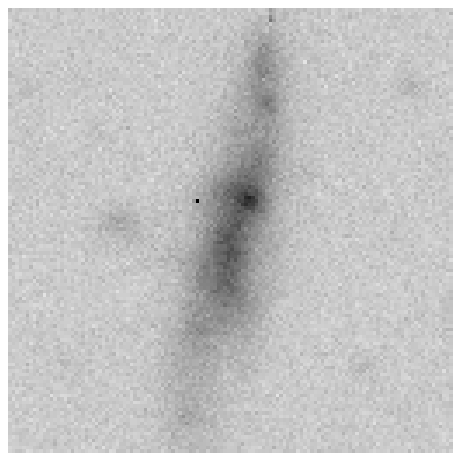,height=3.5cm}}\hspace{1cm}
\subfigure{\psfig{file=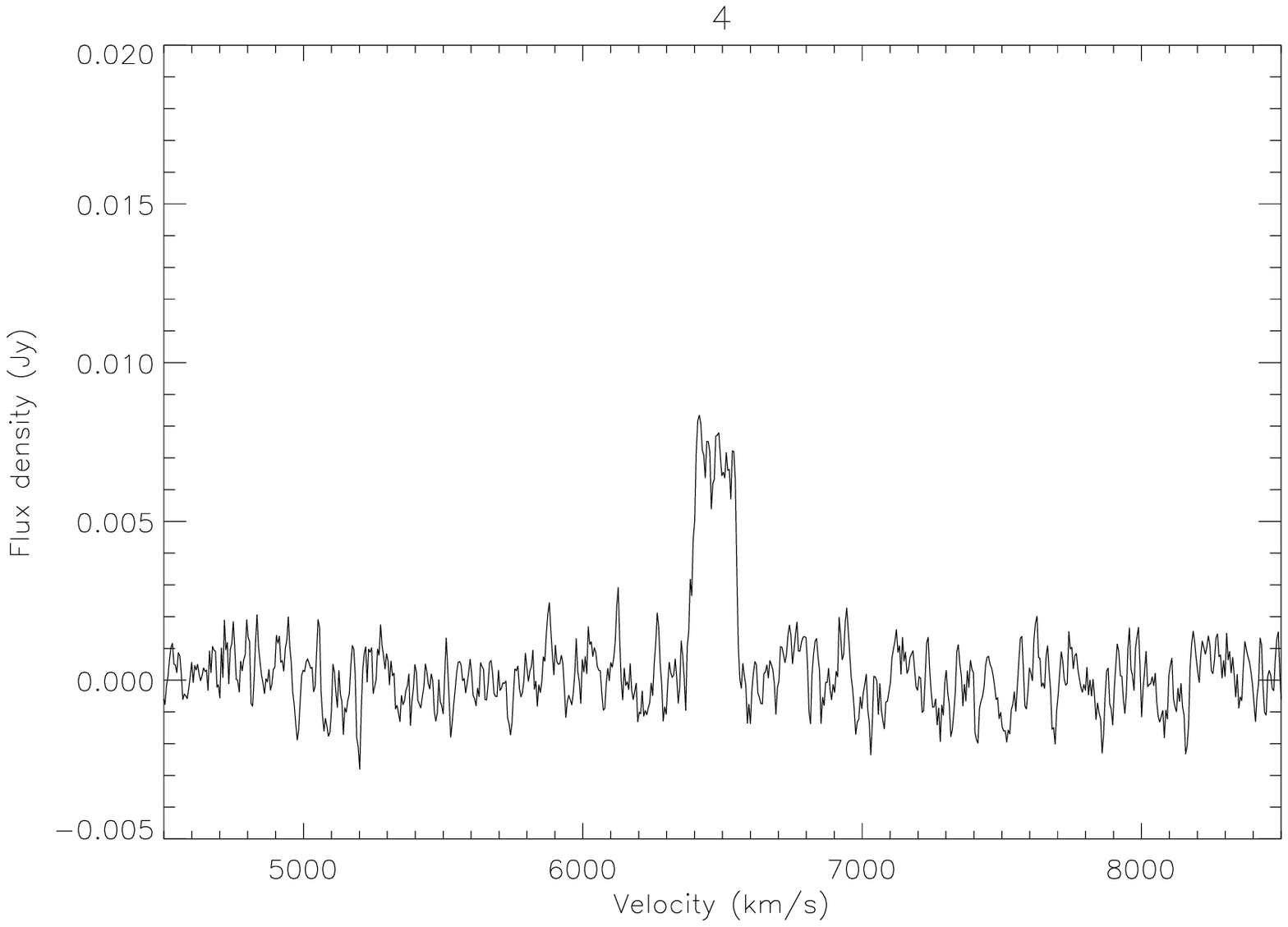,height=3.5cm}} \\
\subfigure{\psfig{file=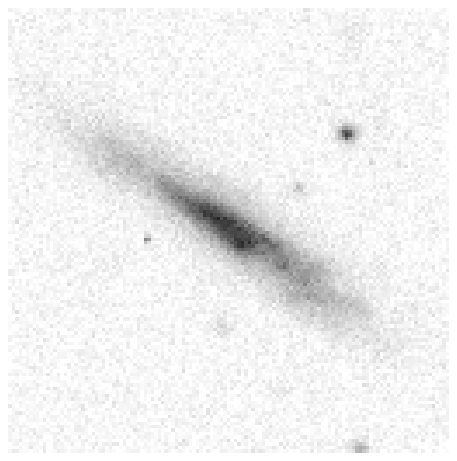,height=3.5cm}} \hspace{1cm}
\subfigure{\psfig{file=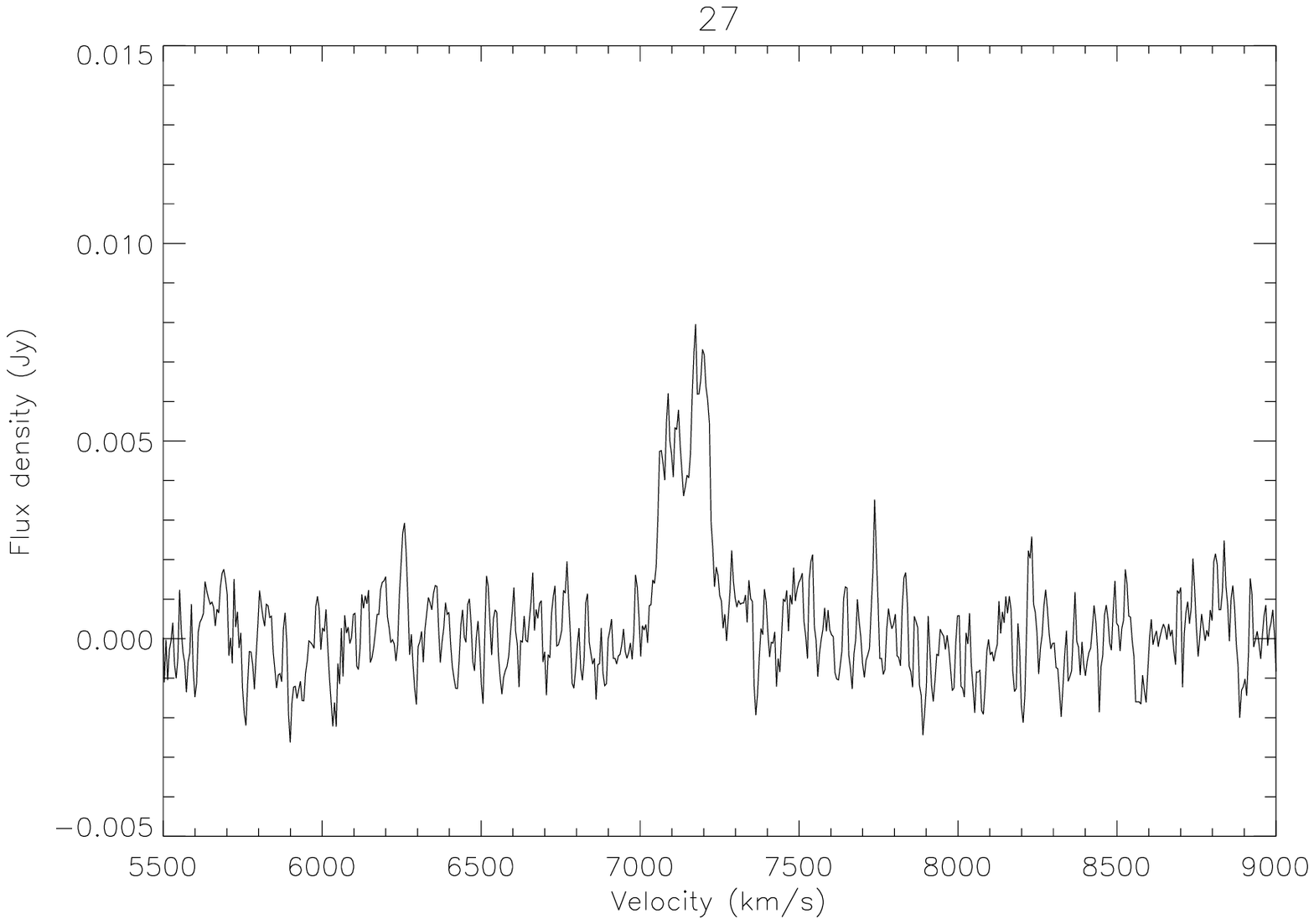,height=3.5cm}} \\
%\subfigure{\psfig{file=t215_plot.ps,width=6cm}}
\caption{\footnotesize{B band images and \HI\, spectra for the 5
non-confused \HI\, detections; respectively nos 144, 158, 249, 4,
27 from our catalogue.}} \label{fig:hispectra}
\end{figure*}

\subsection{Results} \label{sec:hiresults} % 4

Out of the 115 objects we observed in \HI\, just 5 were detected,
3 of which are members (i.e. have line centre velocities in the
range 500 to 2500 \kms\ ). Optical B band images from our deep CCD
frames and \HI\ spectra of these sources are shown in Fig
\ref{fig:hispectra} and a discussion on individual objects is
given in next section. Table \ref{table:dettab} gives their global
optical properties measured in the $B$ band as well as the
integrated line fluxes of the detections: $I_{HI}$, the $W_{50}$
and $W_{20}$ line widths, the centre velocities $V_{HI}$, the
total \HI\ mass $M_{HI} $ and the \HI\ mass-to-light ratio \MHILB\
. The total \HI\ mass, given in \Msun, is derived from the total
flux according to the formula:

\begin{equation}
M_{HI}=2.365 \times 10^{5}\times D^2  \int{S(v)dv}
\end{equation}

\noindent where the distance $D$ is given in Mpc, the flux $S(v)$
in Jy and the velocity $v$ in \kms (Roberts, 1968).

All the given radial velocities are heliocentric and expressed
according to the conventional optical definition
($V=c$($\lambda$-$\lambda_0$)/$\lambda_0$). For objects with
radial velocities that lie inside the range occupied by Virgo
Cluster members, $\sim$ -500 to 2500 \kms, a
distance $d$ of 16 Mpc was assumed. For other objects distances were
calculated using radial \HI\ velocities corrected to the Galactic
Standard of Rest, following the procedure given in the RC3, and
assuming a Hubble constant of $H_0$=75~km~s$^{-1}$~Mpc$^{-1}$.

%The \HI\ line detected towards no. 121 is due to confusion with a
%nearby gas-rich galaxy (see next section for more details).
%
%Listed in table \ref{table:nondettab} are some global optical and
%\HI\ properties for the objects that were not detected in \HI. For
%these we estimated a 3$\sigma$ \HI\ mass detection limits for
%flat-topped profiles, assuming them to be dwarf galaxies in the
%Virgo Cluster with line widths of 75 \kms (typical for dwarf
%galaxies).
%
%\begin{equation}
%M_{HI} \leq 2.365 \times 10^{5}\times D^2  \int{S(v) dv}
%\end{equation}

%

%%%%%%%%%%%%%%%%%%%%%%%%%%%%%%%% DISCUSSION %%%%%%%%%%%%%%%%%%%%%%%%%%%%%%%%%%%%%%%%%%%%%
% \section{Discussion} \label{sec:discussion}
%
%%%%%%%%%%%%%%%%%%%%%%%%%%%% NON DETECTIONS %%%%%%%%%%%%%%%%%%%%%%%%%%%%%%%%%%%%%%%%%%%%
\subsubsection{Notes on individual objects}\label{sec:hidet}
%
%{\sf  6 = VCC 2062:} Not observed by us at Arecibo, since it had already
%been detected in \HI, including mapping with the VLA and the WSRT
%(van Driel \& van Woerden 1989; Hoffman et al. 1993, 1996; Cayatte et al. 1994).
%Classified as E  in the VCC it has (HyperLeda)
%$V_{HI}$=1146 \kms , $W_{50}$=64 \kms\ and $I_{HI}$=7.4 \Jykms,
%implying \MHI= 5.0 10$^8$ \Msun\ and \MHILB=x.x \MsunLsunB.
%Located in a long \HI\ tail linked to the peculiar S0 galaxy NGC 4694
%it could well be a Tidal Dwarf Galaxy (Braine et al. 2001); see also
%Conselice et al. (2003).

%{\sf 77  = VCC 1889:} Not observed by us at Arecibo, since it had already
%been detected in \HI\ (Hoffman et al. 1987). Classified as I? in the VCC, it has
%(HyperLeda) $V_{HI}$=4725 \kms, $W_{50}$=62 \kms, $I_{HI}$=0.23 \Jykms,
%implying \MHI=1.6 10$^7$ \Msun\ and \MHILB=x.x \MsunLsunB.

\noindent {\sf  \textbf{144}:} This object is a cluster member,
which we detected at $V_{HI}$=1036 \kms. Its projected distance
from the cluster centre is $\sim 2^{\circ}$. Its extremely low
central surface brightness (26.6 B mag/\sqarcsec) makes it almost
impossible to see on the B band image. With an \HI\, mass-to-light
ratio of 6.2, it is an extreme case of either a very young galaxy
or a galaxy where the conversion of gas into stars has been very
inefficient. The (B-I) colour for this galaxy is not available,
because the noise in the I band image is too high to measure its
flux.
% 144  & &   123841 & 115843 &  & 19.84 & 26.6 &  & 0.77 &  0.47 &  77 & 125 & 1036 & 17 &   0.32  &   6.1     \\

\noindent {\sf  \textbf{158} (= T152):} This object is a cluster
member at a cluster-centric distance of $\sim 5^{\circ}$, which we
detected at $V_{HI}$=1788 \kms. It is an irregularly shaped galaxy
with several clumps in the B band image.
%    & T152 &   125107 & 120339 & dI &       &      & &  0.91 &  1.71 &  63 &  83 & 1788 & 17 &  1.2  &        \\

\noindent {\sf  \textbf{249} (= UGC 8061):} This object is an
irregular LSB VCC galaxy at cluster-centric distance of $\sim
6.5^{\circ}$. Although our centre velocity and line flux are in
agreement with those of Schombert, Pildis \& Eder (1997) and
Huchtmeier et al. (2000), who found $V_{HI}$= 562 \kms\ and
$I_{HI}$= 2.0 \Jykms, our $W_{50}$ of 72 \kms\ is larger than
their 55 \kms. The B band image shows a very irregular galaxy with
several clumps surrounded by diffuse light.
% 249  & U 8061 &   125644 & 115557 &  & 16.95 & 23.7 & 9 & 1.54 &  2.12 &  69 &  89 &  563 & 17 &   1.4  &  1.9   \\

%{\sf  192 = VCC 2006:} Not observed by us at Arecibo, since it had already
%been detected in \HI\ (Huchtmeier \& Richter 1986;  Haynes \& Giovanelli 1986;
%Hoffman et al. 1987, 1989a).
%Classified as S in the VCC, it has (HyperLeda)
%$V_{HI}$=849 \kms , $W_{50}$=59 \kms\ and $I_{HI}$=1.2 \Jykms,
%implying \MHI=8.5 10$^7$ \Msun\ and \MHILB=x.x \MsunLsunB.

%{\sf = T215:} No confusion is expected of our detection at 7532 \kms\ with 15.8 mag Sc
%spiral VCC 729 at \am{7}{9} distance, for which $V_{HI}$= 7639 \kms\ was measured at
%Arecibo (Haynes \& Giovanelli 1986; Hoffman, Lewis \& Salpeter 1995), as it
%is too far away.
%    & T215 &   122446 & 132026 &  dI &       &      & &  1.14 &  0.36 &  74 & 114 & 7532 & 99.7   &  8.4  &        \\

\noindent{\sf \textbf{ 4} (= T225):} This object, also classified
as dI by Trentham et al. (2003), is a background galaxy, which we
detected at $V_{HI}$=6468 \kms. Its $W_{50}$ line width of 174
\kms\ is much larger than what is usually found for dwarf
galaxies. It could resemble the extreme late type field galaxies
discussed by Matthews \& Gallagher (1997)  in terms of its optical
structure and HI line profile.
%    & T225&   124112 & 105559 & dI &       &      & &  0.87 &  1.13 & 175 & 125 & 6468 & 85.5   &  19.5  &        \\

\noindent{\sf \textbf{27}:}  This edge-on object is a background
galaxy, which we detected at $V_{HI}$=7146 \kms.
% 27   &  &  124934 & 121411 &  &  18.25  &  23.3  & 4 & 0.98 &  1.03 & 197 & 297 & 7146 &  94.7   &  21.8  &   3.1     \\

%\noindent {\sf \textbf{121}*:} Our detection at 1424 \kms\ is confused
%by a nearby galaxy: it is actually that of the SBc spiral NGC 4647
%at \am{2}{8} distance, for which an average $V_{HI}$ of 1415 \kms,
%$W_{50}$=172 \kms\ and $I_{HI}$=8.5 \Jykms\ was reported
%(HyperLeda, from 7 publications).
% 121* & &  124338 & 113720 &  & 19.90 & 25.4 & 5 & 0.88 &  0.59 & 160 & 172 & 1424 & 17 &   ---      &    ---    \\

%{\sf 191 = VCC 1905:} Hoffman (1987) reported a non-detection of
%this object, which is classified as dE2/ImIV in the VCC, with an
%rms noise of 1.2 mJy over a velocity search range of -600 to 3000
%\kms. Our non-detection has an rms noise of 0.9 mJy, over the
%range -1000 to 11,000 \kms.

%{\sf 71 = VCC 1558:} Hoffman (1987) reported a non-detection of
%this object, which is classified as ImV? in the VCC, with an rms
%noise of 1.5 mJy over a velocity search range of -600 to 3000
%\kms. Our non-detection has an rms noise of 1.0 mJy, over the
%range -1000 to 11,000 \kms.

\subsubsection{Notes on non-detections} \label{sec:hinondet}

\begin{figure}
\centerline{\psfig{file=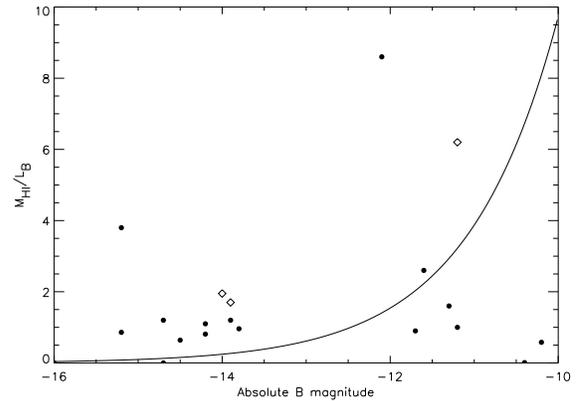,width=8cm}}
\caption{\footnotesize{\MHILB\ 3$\sigma$ detection limit for our
survey as a function of absolute B magnitude of Virgo Cluster
dwarf the galaxies, for an assumed 75 \kms\ line width. Dots in
the plot are dwarf galaxies from the Local Group that have \HI\
emission (Mateo 1998); diamonds are our detections.}}
\label{fig:mtoldetlimit}
\end{figure}

The average noise for our spectra is $\sim$ 1 mJy at a velocity
resolution of 15 \kms. Assuming a typical full width at half
maximum (FWHM) for the \HI\ profile of the dwarfs to be 75 \kms
and a $3\sigma$ detection threshold, at a distance of 16 Mpc, the
non-detections have an average upper mass limit of $\sim 1.5\times
10^7$ \Msun\, which translates into a column density limit of
$\sim 5 \times 10^{18}$ atoms cm$^{-2}$ (assuming a limiting mass
galaxy filling the beam at the cluster distance).
\\
The estimated upper limits of the $(M_{HI}/L_{B})_{\odot}$ ratio
range from 0.1 to 12, depending on the galaxy absolute magnitude.
In Fig \ref{fig:mtoldetlimit} we show this detection limit as a
function of absolute B magnitude (solid line): values of \MHILB\
below the curve are not detectable in our \HI\ observations.
Diamonds in the figure are detections from our sample and dots are
Local Group dwarf galaxies with measured \HI\ (Mateo, 1998):
62$\%$ of these lie above our \HI\ detection limit. If we consider
only the ones that would satisfy our optical selection criteria at
the Virgo distance, the expected detection rate in \HI\ would
still be 50$\%$, for galaxies with the same properties as the
Local Group ones. This is considerably higher than the 4$\%$
detection rate of the observations of our sample and is again a
strong indication of how different the properties of dwarf
galaxies in these two environments are. The plot also clearly
shows that very faint galaxies need high $M_{HI}/L_{B}$ values in
order to be detected.

%This places
%significantly upper limits for the \HI\ mass content of these galaxies. For comparisons
The \HI\ content of dwarf galaxies in the Local Group clearly
distinguishes them as dI (that show \HI\, emission) and dSph/dE
that mostly do not have any detectable \HI\ emission (Mateo 1998).
%Exception to this is Sculptor that has a very low emission in the
%optical centre of the galaxy and a considerably higher emission
%from 2 external lobes $\sim 20$ arcmin far apart from the centre
%(Carignan et al, 1998). Objects of this kind at the Virgo
%distance would be completely included in the Arecibo beam.
In our sample, among the 99 objects with given morphologies that
we observed in \HI\, we find the following distribution: 31 dE, 43
dI, 20 VLSB and 5 not classified. Regardless of their
morphological type our results are consistent (even if not
conclusive) with the idea that the gas in these galaxies has been
either stripped away or efficiently converted into stars. The
results from the \HI\ observations are also consistent with those
for the average value of the B-I colours: on average, the galaxies
in our sample do not show signatures of a young stellar
population, suggesting that the cluster environment has accelerated their evolution.

A pilot sample of LSB dwarf galaxies identified in the Millennium
Galaxy Strip (MGS) was also observed during this Arecibo run. As
already pointed out, this is a dataset in the field that is
identical to ours in the Virgo Cluster and the same detection
algorithm was applied to it, in order to obtain a catalogue of
dwarf LSB galaxies (Roberts et al, 2004). The \HI\ detection rate
for this pilot sample is considerably higher than for our sample
in the Virgo Cluster: 4 out of the 14 ($\sim 30 \%$) observed MGS
field galaxies show \HI\ emission, compared to 5 out of 115 ($\sim
4\%$) in the cluster. Consistent with the well-estabilished trend
for clusters to host gas-poor dwarfs, this result confirms the
importance of the role played by the environment in the evolution
of the gas content of dwarf galaxies.

As a last remark, our \HI\ observations of the Virgo Cluster
sample are also important because they rule out the possibility
that our catalogue is highly contaminated by gas-rich background
or foreground galaxies - this being one of the main worries when
discussing the faint-end-slope of the Luminosity Function of a
cluster or its dwarf galaxy content. According to our
observations, the non-detections are also
non-background/foreground (\HI\ -rich) galaxies in the velocity
range of -1000 to 11000 \kms.

%In figure \ref{fig:himasslimit} we plot the \HI\ mass detection limit for
%a standard spiral galaxy (with rotational velocity width of 200 \kms) as a function of
%distance in the velocity range that we investigated, considering an average rms noise for our
%spectra of 1 mJy.
%\begin{figure}
%{\psfig{file=himasslimit_bg.ps,width=7 cm}}
%\caption{\HI\ mass detection limit for a standard spiral galaxy as a function
%of distance. Here we assume a top-flat emission with a rotational velocity width of 200 km s$^-1$.}
%\label{fig:himasslimit}
%\end{figure}

%\begin{table*}
%\caption{Global properties of the objects of the catalogue without an HI detection.}\label{tab:nondettab}
%\end{table}
% VCC matchtwolists made at 20 arcsec searadius !!!!!!!!!!!!!!!!!!!!!!!!!!!!!!

\section{Discussion}\label{sec:discussion}
Before discussing our results, let us summarise them for clarity:
\begin{enumerate}
\item There are cluster dE galaxies that have a small range of
colours that are preferentially found towards the centre of the
cluster and are gas poor. Dwarf ellipticals are usually passively
evolving (no or little star formation) galaxies and the average
colours that we found are consistent with this picture and
possibly indicative of an older stellar population. \item There
are cluster dI galaxies that have a wider range of colours which
tend to reside in the outskirts of the cluster. They appear to
have star formation regions and the wide range of colours may
reflect different current star formation states.
\item There seems
to be an environmental dependency of the B-I colour distribution
of dwarf galaxies in the Local Universe.
\item There are about 5
times as many dwarf galaxies $(-14<M_{B}<-10)$ per giant galaxy
$(M_{B}<-19)$ in the Virgo cluster than in the general field and
in the Local Group (paper I). A considerable part of these
galaxies seem to constitute a cluster dwarf population rather than
being associated with the giants. Also, the number of dwarfs
possibly associated with giant galaxies in Virgo (i.e. their
projected distance is within the giant tidal radius) is on average
$\sim 13$, compared with $\sim 4$ for the Local Group. Because of
this excess of dwarfs, the cluster cannot have been simply
constructed by infalling small groups like the Local
Group.
\item On the whole, the cluster dwarf galaxy population is
gas poor compared to the dwarf galaxies in the Local Group and in
the field.
\item The luminosity function of the cluster over the
range $(-14<M_{B}<-10)$ is steeper than that in the field. Also,
distinguishing inner and outer regions for the cluster, the LF in
the former region is flatter than in the latter (paper I).
\item
The \HI\ mass function of the cluster appears to be less steep
than the \HI\ Mass Function of the field. Combining this with the observed steep LF leads suggests
that, in the cluster, gas has been more efficiently converted into
stars (Davies et al, 2004); stripping mechanisms might also
however produce this result.
\item In addition, Conselice et al
(2003) have shown that the Virgo Cluster dE population is
dynamically similar to the spiral rather than the elliptical
galaxy population. They argue that this implies that the dE
population is not primordial; rather it is a population that has
fallen into the cluster at a later date.
\end{enumerate}

In the sections below we discuss possible physical processes
acting on the dwarf galaxies in the cluster environment and the
influence that they have on the results described above. In
particular there have been numerous attempts in the past to look
for mechanisms that remove gas from cluster dwarf galaxies to
explain their red colours (no recent star formation) and their
lack of gas. Contrary to this, we believe that our results can be
explained by accelerated star formation that
consumes the galaxy gas, rather than gas stripping mechanisms (see sec \ref{sec:gasloss}). \\
Part of the analysis below is similar to that extensively and
comprehensively discussed in a series of papers by Conselice et al
(2001, 2002, 2003, 2003a). We formulate the problem in  a slightly
different way, concentrating on the mass-to-light ratios required
to avoid gas stripping (see also Davies and Phillipps, 1989). Dwarf galaxies may
in fact be very robust objects in the inter-galactic medium. For example
the recent work by Klenya et al (2002) on the velocity dispersion
of stars in the outskirts of the Local Group galaxy Draco has revised its
mass-to-light ratio from about 100 to 440.
%processes acting on dwarf galaxies in the cluster environment and
%the influence they have on the results described above. In
%particular there have been numerous attempts in the past to look
%for mechanisms that remove gas from galaxies, in order to solve
%the missing dwarfs problem. Contrary to this, we believe that our
%results can be explained by accelerated star formation that
%consumes the galaxy gas, rather than gas stripping mechanisms.

%%%%%%%%%%%%%%%%%% DIFFERENT ENVIRONMENTS %%%%%%%%%%%%%%%%%%%%%%%%%

\subsection{Ram pressure stripping}\label{sec:rampressure}

A cluster like Virgo has a substantial intra-cluster medium (ICM)
which is detected via its x-ray emission (Vollmer et al, 2001). A
galaxy moving through the ICM is subject to a ram pressure that
can possibly strip its gas away, if the ICM pressure on the galaxy
is stronger than its internal gravitational force (Gunn \& Gott,
1972). Ram pressure stripping has often been used in order to
explain the \HI\ deficiency of galaxies in clusters compared with
galaxies in the field (Chamaraux, Balkowski \& Gerard, 1980; van
Gorkom 2003; Lee, McCall \& Richer, 2003).
%The process depends on 3 main parameters: the mass of the galaxy,
%its velocity through the ICM and the density of the ICM.

Following Davies \& Phillips (1989), we can set a limit for the
dynamical mass-to-light ratio $M_{dyn}/L_{B}$ that a galaxy
requires to survive the ram pressure stripping:

\begin{equation}
\label{eq:mtol} \left(\frac{M_{dyn}}{L_{B}}\right) >
\rho_{0}\left(1+\frac{r^{2}}{r_{c}^{2}}\right)^{-(3/2)\beta}
\times \frac{ v_{\bot}^{2}}{\sigma_{gal}} \times
10^{0.4(\Sigma-26.8)}
\end{equation}

\noindent where $v_{\bot}$ is the orthogonal velocity of the
galaxy through the ICM (km/s), $\sigma_{gal}$ is the gas surface
density in the galaxy (\Msun\ pc$^{-2}$), $\Sigma$ is the average
surface brightness of the galaxy (B mag/\sqarcsec) and the density
of the ICM (atom cm$^{-3}$) is a function of position in the
cluster and was parametrized according to a $\beta$-model
(Cavaliere \& Fusco-Femiano 1976).
\begin{figure}
\centerline{\psfig{file=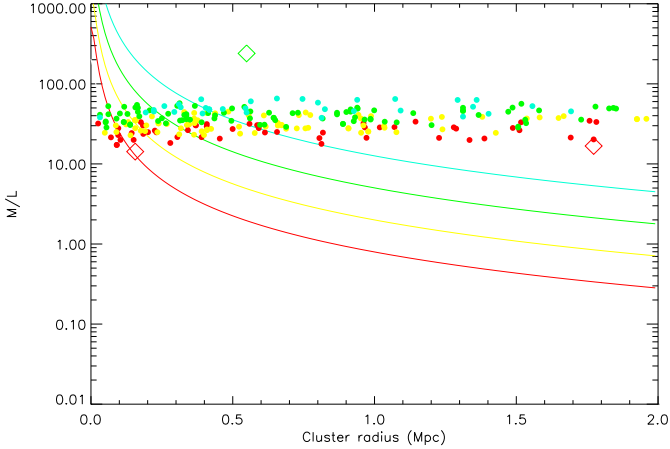,width=9cm}}
\caption{\footnotesize{Curves of the M/L$_{B}$ ratio limit for ram
pressure stripping to occur, as a function of distance from the
centre of the cluster. Galaxies with M/L$_{B}$ above the curves
are able to retain their gas, having a gravitational force
stronger than the ram pressure of the ICM. Different colours refer
to different average surface brightness values in eq
\ref{eq:mtol}: red for $\mu_{0}$=24; yellow for $\mu_{0}$=25;
green for
 $\mu_{0}$=26, blue for has $\mu_{0}$=27. Dots are all
the galaxies from our sample and open diamonds are the 3 cluster
galaxies that have been detected in \HI\ . The dots' colours are
related to the galaxy average surface brightness according to the
lines' colours.}} \label{fig:masstolight}
\end{figure}
For the Virgo Cluster $\beta$=0.45, $\rho_{0}=4\times 10^{-2}$
cm$^{-3}$ and $r_{c}$=13.4 kpc (Vollmer et al. 2001). Also,
assuming $v_{\bot}$=700 km/s (i.e. the average velocity dispersion
of dwarfs in the Virgo Cluster) and $\sigma_{gal}$= 5 \Msun
pc$^{-2}$ (=$5 \times 10^{20}$ atoms cm$^{-2}$, i.e. an average
value for dwarf galaxies; Blitz \& Robishaw, 2000), we obtain
limits for the mass-to-light ratio for different central surface
brightness values of galaxies as a function of distance from the
cluster centre. We take $\Sigma$ to be the mean surface brightness
over the half light radius ($\Sigma = \mu_{0,B}+1.15$). We plot
these curves in Fig \ref{fig:masstolight}: different colours refer
to different central surface brightness. For each curve and each
colour, only galaxies above the curve are not affected by the
ram-pressure stripping and retain their gas. The open diamonds in
the plot show the galaxies in our sample with \HI\ detection and
the dots indicate the non-detections (see sec \ref{sec:hiobs}).

For galaxies without an \HI\ detection the M/L ratio was
calculated using a dark matter halo ($M_{h}$) to baryonic ($M_b$)
mass given in Persic, Salucci \& Stel (1996). Note that
this formula may severely underestimate the mass-to-light ratio
(Klenya et al 2002).
\\
%\begin{equation}
%\frac{M_{h}}{M_{b}}=34.7\mbox{ }M_{b,7}^{-0.29}
%\end{equation}
%\noindent where $M_{b,7}=M_b/10^7 \Msun$ and $M_{dyn} \approx M_{h}$. \\
%Notice that the general trend of increasing dark matter fraction
%with decreasing galaxy mass is indeed consistent with other
%studies of the internal kinematics of dwarf spheroidal galaxies in
%the Local Group for which velocity dispersions of a significant
%number of stars have been derived (Mateo, 1997; Klenya et al,
%2002).
For the 3 galaxies with \HI\, detection, the M/L ratio was
calculated making use of the measured velocity width as follows:

\begin{equation}
M_{dyn} \geq V_{rot}^{2} \times R_{opt}/G
\end{equation}

\noindent and assuming the rotational velocity $V_{rot}$ of the
galaxy to be half of $W_{50}$, and the optical radius $R_{opt}$ to
be 2 disc scale lengths (h):

\begin{equation}\label{eq:mdyn}
M_{dyn} \geq 0.5 \times W_{50}^{2} \times h
\end{equation}

\noindent Note that the values obtained in this way are actually
lower limits, as the \HI\ radius is usually larger than the
optical one. Our calculated values for the 3 galaxies
with \HI\ detections lie above the corresponding $M_{dyn}$/$L_{B}$
limit for the occurrence of ram pressure stripping.
%according to the M/L-luminosity relationship given in Mateo 1994 for early-type dwarf galaxies of the
%local group; different colours for the dots refer to different central surface brightnesses of 1 mag
%range.
%For each curve and colour, galaxies above the curve survive the ram-pressure stripping and
%galaxies below don't.
%Before discussing the plot, we should underline that the Mateo relationship
%assumed here is probably an underestimation of the M/L ratio, as he writes in his paper (higher
%M/L ratios have been found for example  for Draco when extending the measurement of the
%velocity dispersion to stars at a larger distance from the centre of the galaxy; Klenya 2002); furthermore
%we use it also for dwarf irregular, that usually have higher M/L ratios than early-type dwarfs.
%The uncertainties due to these problems would push the points up in the plot, reducing the number of
%galaxies likely to be stripped. \\

According to the lines of different surface brightness in Fig
\ref{fig:masstolight}, we can see that the majority of galaxies
are not likely to be currently affected by ram pressure stripping.
In fact, regardless of their central surface brightness, the plot
shows that galaxies are subject to stripping only if they lie
within the core radius of the cluster, which corresponds to
$R{c}$=0.5 Mpc for the Virgo Cluster. This is in agreement with
the observed gas deficiency of bright galaxies in the centre of
the cluster (see for example Solanes et al (2001) for spiral
galaxies in the Virgo Cluster). Note that the cluster gas density
falls rapidly to $2.8 \times 10^{-4}$ cm$^{-2}$ at the core radius
of 0.5 Mpc  - thus over most of the cluster the gas density is
much lower than that assumed in simulations of ram pressure
stripping (Marcolini et al., 2003). We should also take into
account that the positions plotted for our galaxies are projected
ones: galaxies at a small projected distance to the centre may
actually lie in the outskirts of the cluster. The actual distance
of such a galaxy would then be larger and the point representing
it would move to the right in our plot, thus increasing the number
of dwarfs that are not affected by ram-pressure stripping.

%Also, Vollmer et al. (2001) show that the amount of possible
%stripping and thus the \HI\ deficiency depends strongly on the
%galaxy orbit. Critical parameters are: minimum distance to the
%cluster centre, maximum velocity and inclination of the disk with
%respect to the orbital plane. Different combinations of these
%factors result in very different effects: the galaxy can survive
%the stripping; the interaction with the ICM can completely deplete
%it of its gas; gas can be re-accreted and star formation can be
%triggered, resulting afterwards in a depletion of the \HI\
%content.
%Vollmer et al. also show
%that even if stripped of their gas,
%for a maximum ram pressure of 1000-5000 cm$^{-3}$ (km/s)$^{2}$ and low inclination angle ($<$ 45) galaxies
%can re-accrete their gas. Finally for a maximum ram pressure of  2000 cm$^{-3}$ (km/s)$^{2}$
%and small inclination angle ($<$20) the SF in central regions of the galaxy can increase due to
%increase in the central surface density of the gas. This would accelerate SF and produce again a depletion
%in \HI\ content. \\

%Nevertheless, all of the above mentioned processes require the
%galaxy to approach the cluster centre to within about a cluster
%core radius, so they have not necessarily affected all the cluster
%members.
In fact, it is possible that none of our detected dwarfs are
actually in the cluster core at all. Dwarf galaxies within the
core are subject to tidal forces that could pull them apart. We
can estimate the minimum size ($r_{min}$) of a dwarf galaxy that
would not be tidally disrupted at the cluster core (that is the
distance at which the tidal stress is maximum; Merritt 1984),
using the definition of Roche limit:
% \footnote{The Roche limit is the
%minimum distance from a primary body at which gravitational force
%alone can hold together an orbiting satellite}:

\begin{equation}
r_{min}=R_{c}\left(\frac{M_{dwarf}}{3M_{cluster}}\right) ^{1/3}
\end{equation}

\noindent where $R_{c}$ is the cluster core radius (0.5 Mpc;
Binggeli, Tammann \& Sandage 1987),
 $M_{dwarf}$ is the mass of the dwarf
($\approx10^{7}$ $M_{\odot}$) and $M_{cluster}$ is the mass within
the core radius ($\approx10^{14}$ $M_{\odot}$). This leads to a
size of order 3 kpc, very similar to the sizes of our dwarf
galaxies (1 kpc $\approx$ 10 arcsec at the distance of the Virgo
cluster), see also Merritt (1984). We conclude that dwarf galaxies
passing through the core will be severely disrupted by tidal
forces (cf intra-cluster light, stars and planetary nebulae; Gregg
\& West, 1998; Arnaboldi et al 2002). Given that most of our
galaxies in the projected core region are mainly spherical dEs
(Fig \ref{fig:ratiodEdI}), we conclude that it is unlikely that
they are actually in the core (see also Ichikawa et al. 1988, AJ,
96, 62 ; Carey et al, 1996; Secker et al (1998); Adami et al.,
2000; Oh \& Lin, 2000; Conselice et al, 2001; Gnedin 2003).

Dwarf galaxies are resilient in the cluster because they have high values of M/L, but the ICM ram pressure
may be important in triggering starbursts. Numerical simulations
carried out with smoothed-particle hydrodynamics (Abadi, Moore \&
Bower 1999; Shulz \& Struck 2001 ) and with N-body codes (Vollmer
et al 2001) show that, for small inclination angles (nearly
edge-on) the ram pressure leads to a temporary increase of the
central gas surface density. This, in turn, may give rise to an
episode of star formation. We will consider the gas consumption
through induced starbursts in the last section of this chapter.
%This leads to the other possible explanation that we want to
%consider for the \HI\ 'deficiency' and B-I colour distribution
%found in our sample of galaxies:

%The question arises why, although according to their position in
%the cluster the majority of our galaxies should not have been
%stripped, our \HI\ observations are consistent with them being
%\HI\ poor - down to a detection limit of $\sim 1.5 \times 10^{7}$
%\Msun.

\subsection{Supernovae driven winds}\label{sec:sn}

In almost all CDM models, supernova driven winds play a crucial
role in suppressing the formation of stars in small dark matter
halos (dwarf galaxies), since they could, in principle, completely
blow away its gas. As stated in the introduction this cannot be
the complete solution to the sub-structure problem, because of the
large numbers of dwarfs found in some, but not in other,
environments.

For a supernova driven wind to expel the gas from a dwarf galaxy,
the energy in the wind has to exceed the gravitational binding
energy (Dekel \& Silk, 1986). Davies \& Phillipps (1989)
showed that these energies are about equal for a dwarf galaxy with M/L of 100,
leading to some uncertainty in whether this is
a viable method at all for removing gas from dwarf galaxies.
Following Davies \& Phillipps 1989, we can write this condition
as:

\begin{equation}
T_{W}>10^{3} (M_{dyn}/L) r^{2} 10^{0.4(26.8-\Sigma)}
\end{equation}

\noindent where $T_{W}$ is the wind temperature and $r$ is the
galaxy radius in kpc. For $\Sigma=24.5$ B mag/\sqarcsec\ and $r=1$
kpc, $T_{W}>10^{4}(M_{dyn}/L)$ for gas stripping to occur. Dekel
and Silk (1986) give the range of $T_{W}$ as $6 \times 10^{4}$ to
$6 \times 10^{5}$ (Davies
\& Phillipps, 1989).
This uncertainty is further compounded by the
confining pressure of the ICM for cluster galaxies (Babul \& Rees,
1992) and supported by the recent evidence of the large value of
the $M_{dyn}/L$ for dwarf galaxies in the Local Group (Klenya et
al, 2002).

The above conclusion is also further supported by recent, more
detailed simulations, by Mac Low \& Ferrara (1999) who investigated
starbursting dwarf galaxies with baryonic masses $M_b=10^6-10^9$
\Msun\ and supernova (SN) rates from one per $3\times 10^{4}$ yr
to one per $3\times 10^{6}$ yr. They showed that the critical
masses for which the whole gas content is blown away are very low
($\leq 10^{6}$ \Msun): in almost all the cases,
the starburst is not energetic enough and creates only holes in
the regions surrounding the supernovae event. Note that the total
baryonic (stellar + gas) mass that we expect for our galaxies lies
within the range they investigated: the stellar masses for our
galaxies are $\sim 1.6 \times 10^6$ to $\sim 6 \times 10^7$ \Msun\
(since their magnitude range is $M_B \approx -10$ to -14); our
\HI\ observations (see sec \ref{sec:hinondet}) indicate that they
have \HI\ masses below about $1.5 \times 10^7$ \Msun. The SN rates
explored by them are also consistent with the starbursts that we
consider. These results, along with the observed high values of M/L,
must give concern as to whether SN driven winds are a viable gas
stripping mechanism (see also Mayer and Moore 2003).

\subsection{Tidal interactions}

Gravitational interactions can cause mergers, reshape the galaxy
and its content or trigger starbursts. Contrary to ram pressure
stripping, numerical simulations show that they may be important
even at large distances from the cluster centre and that the
numerous high-speed encounters occurring in clusters can give rise
to morphological transformations, i.e. disc systems into dwarf
Spheroidals and dEs (Moore et al,, 1999; Moore 2003).

%Gnedin (2003) argues that tidal interactions with neighbouring
%galaxies and with the dark cluster halo are the most favoured
%processes to explain the morphological chemical and dynamical
%evolution of galaxies in clusters.
Tidal interactions in the cluster could have produced in the past
instabilities that accelerated the star formation activity of the
galaxies, resulting in the currently observed  B-I
colours\footnote{Here, again, we are assuming that all the
galaxies in our sample have similar metallicity, so that redder
colours means older stellar population.} and depletion of the atomic gas. This possibility is discussed in
the following section.

The colour distributions in figure \ref{fig:histo_colours} and the
dynamical timescales in table \ref{table:colours} indicate a
dependence of galaxy properties with the crossing time. The
crossing time is related to the density of the cluster (not its
total mass), with the most dense clusters having the shortest
crossing time. This suggests that local interactions of galaxies
have a large influence on their subsequent evolution. The
importance of these interactions depends on the encounter rate,
which is related to  the number density ($n$) of galaxies in the
specific environment considered, the interaction cross-section
($\varrho$), the velocity of the galaxy through the cluster ($v$)
and the age of the cluster ($t$). Using a simple rate argument we
can write the characteristic number of interactions per galaxy as:

\begin{equation}\label{eq:nrate}
N\approx n\varrho v t
\end{equation}

%\noindent For a cluster age of $\sim 10$ Gyr, we can write:
%\begin{equation} \label{eq:nrate}
%N\approx 4\left(\frac{n}{250 \mbox{ Mpc$^{-3}$}}\right)\left(\frac{r_{p}}{20\mbox{ kpc}}\right)^{2}
%                 \left(\frac{\sigma_{v}}{1000 \mbox{ km s$^{-1}$}}\right)
%\end{equation}
%\noindent where $v=\sqrt{2}*\sigma_{v}$ (Mihos, 2003). \\
%This gives and average encounter rate of $\sim$ 1 for Virgo (just???),
%and for comparison with the rate scales with the ratio of number density and velocity
%dispersion of the different environments to Virgo ones, this would give a rate of ?? for Ursa and
%of ?? for Fornax.
\noindent If, for the sake of simplicity, we assume the cluster
age to be the same for the different environments that we
considered in sec \ref{sec:diffenv}, then we can calculate the
relative number of encounters compared to that occurring in the
Virgo Cluster as a function of $ n$ and $\sigma_v$ (=$v/\sqrt{2}$)
only, the cross section being the same. Using, for consistency,
the information we need in eq \ref{eq:nrate} for the 3 clusters
(the total number of galaxies, virial radius and velocity
dispersion) given by Tully et al. (1996), we find that the number
of encounters experienced by galaxies in the Virgo Cluster is 3
times lower than in Fornax and 20 times higher than in the Ursa
Major cluster \footnote{Note that introducing a dependence on the
estimated cluster ages, does not alter this result.}. If
encounters promote star formation, this would have lead to an
increased past star formation activity going from the Ursa Major
to the Virgo and Fornax Clusters and could thus account for the
reddening of the colour distribution.

The rate of interaction is obviously density dependent
and the balance between the destruction vs. production rates is
delicate, as noted for example in the Coma Cluster by Adami et al.
(2000). Conselice (2002) show that a steepening or a flattening of
the faint-end-slope of the evolved LF in a cluster is function of
the maximum number of tidal interactions experienced by the
galaxies. These simulations show that there should be a density
threshold above which destruction is more efficient than creation.
This argument can also be easily used to interpret the difference
in the inner and outer Luminosity Functions of the Virgo Cluster
that we showed in paper I. The dwarfs that now remain in the
cluster would then be 'survivors' to this processes and they would
have continued to form stars over a longer period than systems
that were destroyed (for example in the cluster core).

% Results from the VCC and subsequent studies based on it, seem to
%indicate that the dwarf population of the Virgo cluster consists
%of at least two different components: an old gas-poor one and a
%lately accreted one that shows clear signs of un-relaxation
%(Conselice, Gallagher \& Wyse, 2001; Conselice et al. 2003). {CUT
%THIS?!? \it This recent analysis by Conselice et al, seems to
%favour the harassment scenario (Moore, 1999) as the likely origin
%of the latter population, suggesting also the possibility of the
%existence of extremely faint dEs that would be the ultimate
%product of the evolution of dIs that were accreted into clusters
%over many epochs, along with their spiral companions.}
%A further prediction of the harassment model is that dEs should be
%embedded within very low surface brightness ($\geq 27 $
%mag/\sqarcsec) tidal streams of stellar debris and extremely
%compact stripped nuclei. The surface brightness of these streams,
%however, is predicted to be below the sky noise of our present
%survey data. Evidence in support of such predictions has been
%found in a small number of diffuse light trails (Gregg \& West,
%1998), in intra-cluster stars (Arnaboldi et al, 2002) and also by
%the discovery of a new population of galaxies in the Fornax
%Cluster constituted by ultra-compact dwarfs (Drinkwater et al,
%2003).

To complete this picture, we can also investigate how likely it is
that a dwarf galaxy will have an interaction with another cluster
galaxy capable of completely disrupting it and/or stripping its
gas away. To estimate this we can calculate the interaction cross
section, $\varrho$ ($=\pi r_{p}^{2}$, with $r_{p}$ impact
parameter), as the square of the Roche limit distance, $R$:

\begin{equation}
\varrho \approx R^2 \approx r^{2}(3M_{gal}/M_{dwarf})^{2/3}
\end{equation}

\noindent where $M_{gal}$ is the mass of the galaxy the dwarf is
interacting with and $r$ is the dwarf galaxy radius. For a given
cluster the number of interactions depends only on the
cross-section $\varrho$. It is straightforward to show that a
$10^{10}$ $M_{\odot}$ galaxy is about 45 times more likely to
interact in this way with another $10^{10}$ $M_{\odot}$ galaxy
than it is with a $10^{7}$ $M_{\odot}$ dwarf galaxy. The reason is
that the dwarf galaxy, because of its small size has to be very
close to the large galaxy ($\lesssim 10 $ Kpc) for severe tidal
disruption to occur (e.g. the Sagittarius dwarf galaxy of the
Local Group). This would be a problem if most of the Virgo dwarf
galaxy population were closely associated with the giant galaxy
population. In Fig \ref{fig:DGR_cerchi}, however, we have shown
that there is a population of Virgo dwarf galaxies that is not
associated with the giants. As we do not see large numbers of
large galaxies undergoing disruptive tidal interactions we have to
conclude that dwarf galaxy disruptive interactions are also
uncommon. This also does not appear to be a viable method of gas loss.

If the cluster dwarf galaxy population originated from a tidally
truncated population of larger field galaxies then their tidal
radii should be of order
$R_{c}\frac{\sigma_{dwarf}}{\sigma_{clust}}$ where
$\sigma_{dwarf}$ ($\lesssim 10$ km s$^{-1}$) and $\sigma_{clust}$
($ \approx 700$ km s$^{-1}$) are the dwarf galaxy and cluster
velocity dispersion respectively. The smallest calculated radius
is of order 7 Kpc; this is larger than the dwarf galaxies in our
sample ($\approx 1$ kpc) implying that they must sit in much
larger dark matter halos. Moore et al 1998 also give a prediction
for values of harassed galaxies' effective radii; these range from
5.3 to 1.6 Kpc, again larger than the typical values for galaxies
in our sample. This poses a problem to the harassment scenario as
a means of producing the dwarf galaxies in our sample through
morphological transformation of large discs. The transformation of
infalling gas rich (already small) dwarf irregulars, like those
detected by us, is however still quite possible.

\subsection{Gas loss through enhanced star formation}\label{sec:gasloss}

Direct evidence for galaxy evolution in clusters is given by the
Butcher \& Oemler effect (BOE; Butcher \& Oemler, 1978). Three
triggering mechanisms have been suggested for the BOE: ram
pressure by the hot intra-cluster medium (Dressler \& Gunn 1990),
galaxy-galaxy interactions (Lavery \& Henry 1988) and tidal
triggering by the cluster potential (Henriksen \& Byrd 1996). The
lack of correlation of the BOE with the X-ray luminosity, however,
rules out the possibility that the former mechanism is the unique
one at work. On the other hand, several studies of both
statistical samples and individual interacting systems have shown
that the optical and infrared colours of peculiar galaxies can be
explained in terms of bursts of star formation triggered by tidal
interactions (Larson \& Tinsley, 1978; Young et al, 1986; Larson
1986; Liu \& Kennicutt, 1995; Barton Geller \& Kenyon, 2000,
2003). Theoretical work employing smoothed particle hydrodynamics
(SPH) also gives support to the notion that galaxy interactions
can drive significant inflows of gas under a wide range of
conditions and rise the star formation rate by more than an order
of magnitude (Noguchi \& Ishibashi, 1986; Barnes \& Hernquist,
1995; Mihos \& Hernquist 1996).

Our view is that our observations can be explained not by gas
stripping mechanisms, as described above, but by accelerate star
formation in infalling harassed dI galaxies or dark haloes that
have no stars - the gas that once was there is still there, but
now in the form of stars. This is witnessed by the very different
cluster and field Luminosity and \HI\ Mass Functions, the spatial
distribution of dE and dI galaxies, their very different
morphologies and gas content. We suggest that cluster dwarf
galaxies have undergone accelerated evolution compared to the
field.

Although we have shown that tidal interactions that strip a dwarf
galaxy of its gas must be rare, numerical simulations do show that
the several high-speed encounters that disc systems undergo in
clusters can transform infalling galaxies into dEs. The
continually varying tidal force compresses the gas, promoting
enhanced star formation compared to dI galaxies in the field. As we have
suggested in the previous section we believe that the galaxies
that undergo this transformation are the gas rich dIs observed on
the cluster edge.

Can enhanced star formation alone explain the colour distribution
of our galaxies and the possible lack of \HI\ emission from the
dwarfs of our sample? If we assume the initial \HI\ mass of a
dwarf to be 10$^{7}$ \Msun\ and that each tidal interaction
triggers a starburst of $\sim 10^{8}$ yr duration and consumes
$\sim 10^{6}$ \Msun\ of gas (Leitherer \& Heckman, 1995), then 10
such starbursts can fully exhaust the \HI\ content of the dwarf
\footnote{Supernovae driven winds produced by an event of this
kind would not blow away the gas of the galaxy, as shown in
section \ref{sec:sn}}. This would make dwarfs in clusters very
different from isolated dwarf irregulars that typically have long
gas depletion timescales of $\sim$ 20 Gyr and which have
experienced (and will continue to experience for at least another
Hubble time) a slow, but constant, star formation activity (van
Zee, 2001).

What would happen to the colours and absolute magnitudes of the
dwarfs in our simple model of several starbursts? Numerical
simulations on the effects of starburst on low surface brightness
galaxies show that surface brightness is virtually unaltered by
these episodes, while the total colours can change significantly
(O'Neil, Bothun \& Schombert, 1998). According to Leitherer \&
Heckman (1995) the change in (B-I) colour in 10$^8$ yr would be
$\sim$ 1 mag, regardless of metallicities and for either
continuous star formation activity or a single starburst. The
colour distribution of our sample is consistent with such a
scenario. The galaxies' average (B-I) colour is $\sim$ 1.7, which
is about the typical colour for metal poor globular clusters
(Reed, 1985).
%Their colours can therefore be accounted for by age only:
%they are a population of galaxies that have undergone different
%episodes of star formation, possibly triggered by the cluster
%environment and have thus consumed their gas content.
We should also note here that the three \HI\ detections in our
sample are extremely blue if compared with the average value of
the colour distribution and with the colours of metal poor
globular clusters: 144 is not visible at all in the I band, 249
and 158 have (B-I) colours of 1.26 and 1.01 respectively,
indicating a younger population. These galaxies might then be
interpreted as being part of an infalling population of initially
gas-rich dIs that are possibly undergoing morphological
transformations that will eventually turn them into gas-poor
galaxies (see also Conselice et al 2003).

There are two further points to consider: the presence of
intra-cluster light and planetary nebulae has been used to argue
for the large scale stripping of material from galaxies in the
cluster. All of the above have only been detected within the Virgo
Cluster core. We argued earlier that galaxies entering the core
would be tidally disrupted and use this to suggest that our sample
galaxies are not in the core, but their positions are projected on
to it. Total tidal disruption is possible within the cluster core.
Secondly the observed luminosity-metallicity relation of dwarf
galaxies has been used to argue that the more massive dwarfs
retain their gas longer and so produce more generations of stars
and have an higher metallicity. Phillipps \& Davies (1989) showed
that this can be also explained as a surface
brightness-metallicity relation, in which there is a gas column
density cut-off for star formation.

\section{Conclusions} \label{sec:conclusions}

We have presented (B-I) colour and sensitive 21cm observations of
a sample of Virgo Cluster dwarf low surface brightness galaxies
that includes previously un-catalogued cluster members. These
follow ups complete our work in paper I on the contribution of
these galaxies to the faint-end-slope of the Luminosity Function.
The \HI\ observations also complete an earlier paper (Davies et
al, 2004) aimed at investigating possible under-evolved \HI\
clouds in the cluster.

We find that, on average, colour trends are weak functions of both
total B magnitude and central surface brightness of the galaxies.
We have also compared the properties of dwarf galaxies in the
cluster with those in other environments: dwarfs follow a similar
density-morphology relation as the brighter galaxies. This is
strange, given their predicted very different histories according
to CDM theories (building blocks as opposed to fully assembled
objects). \\
Cluster dwarfs are generally gas poor and red compared to dwarfs
in the field, but they have average (B-I) colour and a velocity
dispersion more like the spiral galaxy population.\\
%There are dwarfs that are not primordial, in the sense that they
%formed in situ as the cluster formed - their velocity dispersion
%is closer to that of the spiral galaxy population (Conselice et al
%2003).
The dwarfs in our sample do not seem to come from the infall of
units like the Local Group: the dwarf-to-giant ratio is too
high.\\
The cluster luminosity function is steep while the \HI\ mass
function is shallow compared to the field.

Different mechanisms at work in the cluster environments were
discussed and their relative efficiencies compared. We conclude
that enhanced star formation activity due to tidal interactions
might account for the different properties of dwarfs in the
different environments we compared. We believe that the dwarfs in
our sample are not larger galaxies stripped of their gas or
tidally transformed and have not been created in tidal
interactions of larger galaxies.

The picture that we suggest is a scenario where tidal interactions
play a fundamental role in triggering the star formation in
cluster galaxies. We conclude that the dwarfs in our sample
arrived (are arriving) in the cluster as gas rich dwarfs
converting their gas into stars rapidly. It is not clear to us how
this might be accommodated within the CDM model.

\section{Acknowledgements}
We want to thank the staff of Arecibo Observatory, especially
Tapasi Ghosh, Phil Perillat and Chris Salter, for their help with
the observations and data reduction. The Arecibo Observatory is
part of the National Astronomy and Ionosphere Center, which is
operated by Cornell University under a cooperative agreement with
the National Science Foundation. This research has made use of the
GOLD Mine database, operated by the Universita' degli Studi di
Milano-Bicocca, the Lyon-Meudon Extragalactic Database (LEDA),
recently incorporated in HyperLeda, and the NASA/IPAC
Extragalactic Database (NED) which is operated by the Jet
Propulsion Laboratory, California Institute of Technology, under
contract with the National Aeronautics and Space Administration

\label{lastpage}

\end{document}